# Bio-inspired vascularized electrodes for high-performance fast-charging batteries designed by deep learning

[1, 3]Chenxi Sui, [2, 3]Yao-Yu Li, [1]Xiuqiang Li, [1]Genesis Higueros, [1]Keyu Wang, [1]Wanrong Xie, [1]Po-Chun Hsu*

[1]Thomas Lord Department of Mechanical Engineering and Materials Science, Duke University
[2]Department of Chemical Engineering, University of Washington, Seattle
[3]These authors contribute equally
*Corresponding email: pochun.hsu@duke.edu

## Abstract

Slow ionic transport and high voltage drop (IR drop) of homogeneous porous electrodes are the critical causes of severe performance degradation of lithium-ion (Li-ion) batteries under high charging rates. Herein, we demonstrate that a bio-inspired vascularized porous electrode can simultaneously solve these two problems by introducing low tortuous channels and graded porosity. To optimize the vasculature structural parameters, we employ artificial neural networks (ANNs) to accelerate the computation of possible structures with high accuracy. Furthermore, an inverse-design searching library is compiled to find the optimal vascular structures under different industrial fabrication and design criteria. The prototype delivers a customizable package containing optimal geometric parameters and their uncertainty and sensitivity analysis. Finally, the full-vascularized cell shows a 66% improvement of charging capacity than the traditional homogeneous cell under 3.2C current density. This research provides an innovative methodology to solve the fast-charging problem in batteries and broaden the applicability of deep learning algorithm to different scientific or engineering areas.

# 1. Introduction

The development of fast-charging batteries is one of the most essential milestones to push forward the electrification of vehicles and to mitigate the impinging crisis of climate and global warming. [1-2] As identified by the U.S. Department of Energy, extreme fast charging (XFC) electric vehicles (EVs) should be able to recharge in less than 10 minutes to provide 200 miles of driving range. [3] The ambitious objective poses grand challenges and opportunities for scientists and engineers to overcome conventional hurdles at various aspects and length scales. One bottleneck is the tradeoff between areal capacity and charging rate. Most batteries cannot maintain a high areal capacity under a high charging rate, due to the poor utilization of deep electrode materials. [4, 5, 18, 20, 21] Although introducing porosity can be helpful, it has to be achieved without sacrificing the cost or energy density at the cell scale. The challenge to overcome the tradeoff between rate performance and mass loading is because randomly-packed porous electrodes tend to be tortuous, which hinders the ion and electron transport within the electrode. Progress has been made to reduce the electrode tortuosity by creating vertical channels through the electrode matrix. [4-18] However, complexity arises when considering the nonuniform distribution of the local reaction current density (reaction rate) along with the electrode thickness. [19] Generally, the reaction rate is higher near the separator than that of the current collector, which implies gradual porosity (higher porosity near the separator-electrode interface and lower porosity near the current collector) also plays a vital role in high battery rate performance. [20-22] Hence, to achieve fast-charging batteries via porous structure engineering, both porosity distribution and the vertical channel parameters need to be considered, which quickly becomes a highly complex and non-linear design problem.

Nature has already provided plenty of examples to solve this kind of multi-variable transport optimization problem. Billions of years of evolution has created complex transportation system in organisms, such as plant roots, leaf veins, and blood vessels [23-26], in which vascularized channels are evolved to find the optimal balance between mass transport, metabolic efficiency, and uniformity. [24-27]. Inspired by nature, many researchers have designed and manufactured vasculature for efficient transport in applications such as gas sensors, microfluidic systems, and fuel cells. [28-34] Therefore, we hypothesize that such vasculature approach can also be applied in Li-ion batteries to accomplish fast-charging without sacrificing material utilization or capacity, as shown in Figure 1a and parametrized into the numerical model as Figure 1b. Despite having a great potential to outperform traditional methods, the optimization of vascular structures for Li-ion batteries has not yet been well studied, possibly due to the immense parameter space.

Given the physical governing partial differential equation groups for a Li-ion battery cell and the goal of maximizing charging capacity, there are a few possible approaches to determine the optimal vasculature geometry and parameters. The first is to compute all the charging curves for each geometry by finite element method (FEM). This method works for the highly-simplified scenario as in 1D or even some 2D modeling, but it becomes dramatically time-consuming when considering 3D vasculature design. The second method is topology optimization, a method to find the optimal material spatial distribution for specific objective functions and constraints. Topology optimization has been proven to be a powerful approach for design problems in heat exchangers, structural mechanics, [35, 36] and solid oxide fuel cells [37, 38]. Nonetheless, it is nontrivial for lithium-ion batteries to formalize the objective function and implement topology optimization because the parameters of charging capacity computation are inherently time-dependent and the Nernst-Planck equation is nonlinear. Prior reports that used topology optimization to study the battery discharging capacity only solved a 2D problem and used the internal resistance as a surrogate for capacity, which could not obtain a distinctive improvement in the end [39].

Here, we demonstrate an effective and efficient approach to perform both forward- and inverse-design of vasculature battery electrodes based on deep learning. With the rapid development of data science and artificial intelligence, machine learning algorithms have demonstrated their advantages to solve various optimization problems [40], including predicting the optimal operating conditions, charging curves, safety and synthesis methods concerning Li-ion batteries with promising performances. [41-46] To study the influence of vasculature and develop the design tool, we adopt the workflow enabled by the following innovative points:

first, we start with the bio-inspired vascular structures as the initial guesses, which gives the artificial neural network (ANN) a time-saving direction to find the optimal geometry. Second, multiple ANNs with bootstrap aggregating (bagging) ensemble algorithm are used for forward- and inverse-design vascular structure to accelerate the computation efficiency 85 times than traditional FEM. Finally, we apply the as-optimized vasculature idea to both cathode and anode, and the vascular full cells show a 66% capacity improvement compared with homogeneous cell under U.S. Advanced Battery Consortium goal for fast-charging electric vehicles' batteries [1].

## 2. Results and discussion

**Scheme**

The workflow of developing ANNs for vasculature design is illustrated in Figure 1c. The whole process contains the following steps: (1) We created the list of 10 geometric parameters for vasculature (Figure 1b), which is described in Supporting Information Note 1 in detail. (2) For training dataset generation, we randomly chose 4,611 parameter combinations, each of which represents one unique electrode vascular structure. These geometry parameters must also satisfy certain constraints to guarantee the successful formation of the topology structures (e.g., no overlap or penetration). The mass loading and thickness were fixed for all the electrodes to ensure a fair comparison. (3) We fed these geometric parameters into finite element modeling software COMSOL Multiphysics to compute the corresponding charging curves, which will serve as the training dataset for our ANNs. (4) For the deep learning process, we combined the neural networks and bagging ensemble algorithm to improve the prediction stability and the accuracy of our models. The as-trained neural networks were then used to predict all 389,514 charging curves (64,919 possible vascular structures in our geometric parameter space multiplied by 6 charging rates). (5) We demonstrated inverse design capability by compiling the ANN predicted charging curves as the "inverse searching library", which can find the optimal vascular structure under specific targets and limits.

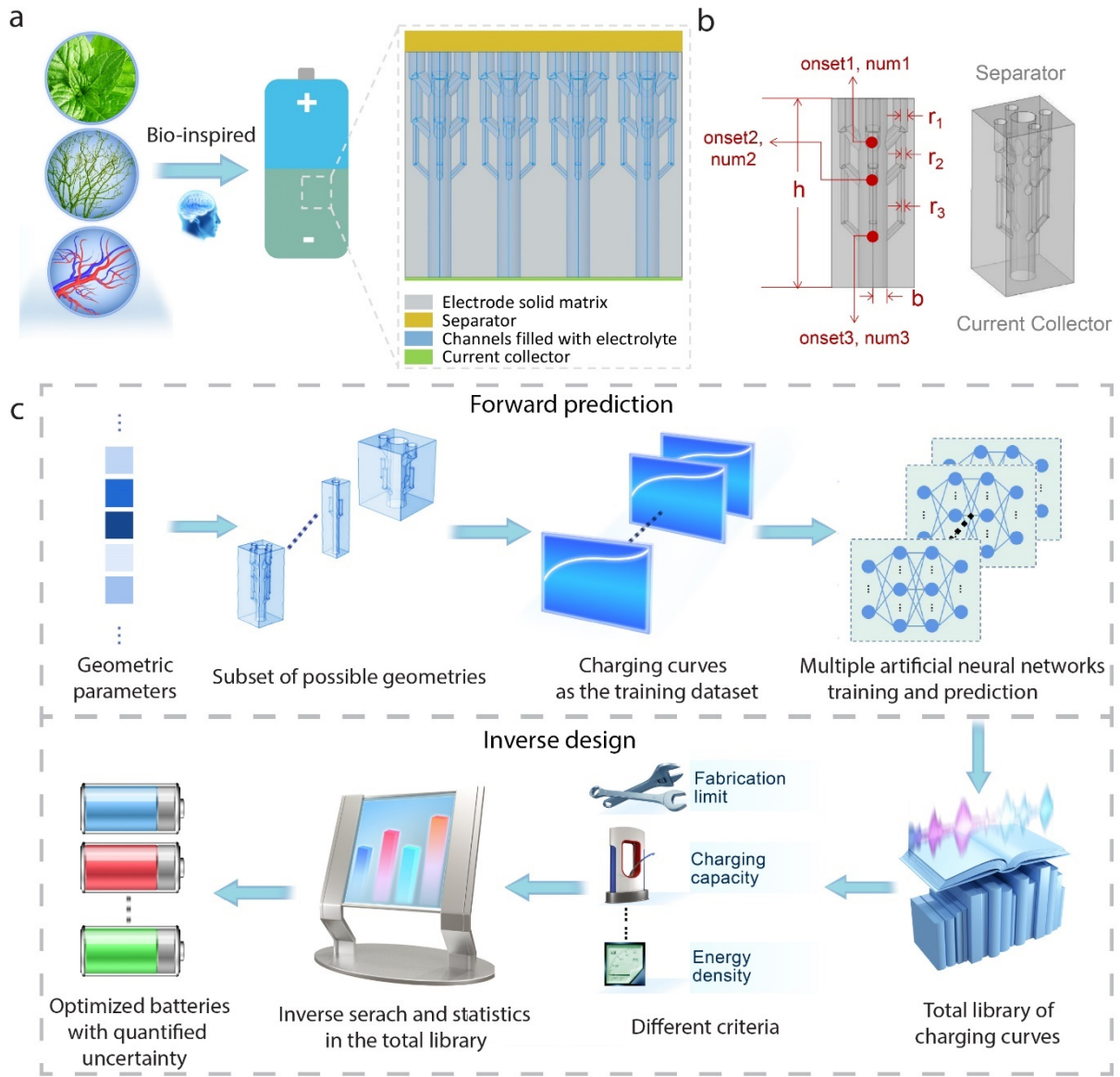

**Figure 1. Schematic demonstration of bio-inspired vascular electrode and workflow of the deep learning optimization process.** (a) The bio-inspired vascular channels for fast-charging batteries. (b) Vascularized electrode geometric model and parameters. The radius of the central channel and the thickness of the electrode are expressed as *b* and *h*, respectively. Three onset points where the branches grow are defined as *onset1-3*. The number of branches on each onset point are labeled as *num1-3*. The radii of branches, *r1-3*, are obtained by multiplying the central channel radius with a ratio *alpha1-3*. That is, *r = b\*alpha*. We assume the radius of branches on the same onset point are equal. (c) The workflow of forward deep learning prediction and inverse design optimization.

**Electrochemical analysis for vascular structures**

Before generating the training dataset and implementing the deep neural networks, we first demonstrate our bio-inspired vasculature idea by choosing specific vascular geometries based on the physical reason of improvement of the vasculature from fundamental electrochemistry and transport theories. We suggest the increased capacity results from the synergistic effect of graded porosity and low tortuosity. The former effect is based on that total the ionic current is larger at the separator side (the boundary condition at the separator/electrode interface dictates all the current is carried by ions), so lowering the transport resistance near the separator can reduce the overpotential or IR drop, which can be achieved by creating a graded porosity that is more porous near the separator, as depicted in Figure 1a. The porosity is defined by the average porosity at a certain xy-plane and is calculated for all four cells we choose here in Figure S19. The latter phenomenon is well understood that low tortuosity can increase the effective ion diffusivity in the porous electrode matrix, thereby mitigating the polarization [4-18]. We summarize the comparison between vascularized, vertical channels, and homogeneous electrodes in the spreadsheet in Figure 2a.

To test the above hypothesis, we defined four different electrodes unit cells in Figure 2b and computed their charging capacity in COMSOL. The geometry models and electrochemical simulation method [47, 48] are described in Supporting Information Notes 1-3. Other geometric factors are: *alpha3* = 0.5, *onset3* = 0.14 mm, $b$ = 0.005 mm. For a fair comparison, we let all four cells have the same thickness and mass loading, and the minimum channel diameter is also the same for all vascularized channeled electrodes. We computed their charging curves under 5C charging rate with the cutoff voltage of 4.3 V versus $Li/Li^+$. As shown in Figure 2c, the capacity is improved by integrating the vertical channels, consistent with previous research on low tortuosity battery electrodes with vertical channels. [4-18] Remarkably, introducing vascular channels further enhances the capacity, which proves our hypothesis that the bio-inspired vasculature can indeed facilitate the electrode design for fast-charging batteries.

For the detailed fundamental electrochemical analysis, we first consider the enhancement by the graded porosity by discussing the penetration depth of Li-ion from the bulk electrolyte (separator side) into the porous electrode. As studied by Newman et al., the electrodes with higher penetration depth are utilized more efficiently. [19] The edge of the unit cell is chosen because it is the farthest place from the channels and represents the most extreme scenario (Figure 2b). If more Li ions can diffuse to this location, the electrolyte supply to the whole cell should be better. We plot the Li ion insertion concentration along the edge of unit cells (defined as the red arrow in Figure 2b) under 10 C charging rate at the cutoff voltage. Note that the top of the unit cell is in contact with the separator (specified in Figure 2b). The concentration decreases dramatically away from the separator due to the depletion of Li ions under high current density (Figure S16). The zoom-in plot between $x = 0.14$ mm and $x = 0.18$ mm (Figure 2d) shows that vascular branches increase the local Li-ion concentration. As seen in Figure 2d, the more branches we have near the separator, the higher Li-ion insertion concentration and reaction penetration depth. Higher Li-ion insertion concentration indicates more Li-ion diffuse into the electrode, and the electrochemical reaction penetrates deeper, explaining the function of the vasculature.

The second factor is the reduced tortuosity, which is characterized by the DC depolarization test. Figure S17 shows the voltage profile of the polarization-depolarization process. The cell is first charged under 0.1C current for 1000 s and then relaxed for 49000 s to reach the equilibrium potential $U(t =\infty)$ ($\Delta U/\Delta t < 0.1$ mV h$^{-1}$). As reported by Li et al., [5] the kinetic process is described as the slope of the logarithm linear fit of $U(t) - U(t =\infty)$, which represent the potential decay factor below:

$$\ln |U(t) - U(t = \infty)| = \text{const} - \frac{t}{t^\delta} \qquad (1)$$

where $1/t^\delta$ is the decay factor. The higher the absolute value of the slope is, the faster the voltage would decay to equilibrium potential, demonstrating an improved kinetic process of the electrode. The curves in Figure 2e show two different kinetic processes. In the beginning (from 1000 s to 2000 s in the blue shaded region), the high slopes represent the fast kinetic process, mainly contributed by the ionic electronic diffusion in the electrolyte liquid phase. Then, after the fast kinetic process, a slow kinetic process follows (red shaded region), which corresponds to the diffusion of Li ions from solid electrode particles to liquid electrolyte. In Figure 2f, we plot the slopes (or voltage decay factors) of different geometric electrodes under these two processes. Vertical channels increase the decay factors from $2.544*10^{-3}$ to $2.643*10^{-3}$ for a fast process and from $8.38*10^{-5}$ to $9.41*10^{-5}$ for a slow process, respectively. The decay factors remain similar values by branching out more vascular channels (2-branch and 4-branch). This makes physical sense because the channels are straight, and their radii are on the same order of magnitude, which suggests similar tortuosity.

The two physical factors listed above provide the vasculature with high C-rate capacity improvement, which can be characterized by the high C-rate pulse charging. The lower voltage during the high C-rate pulse charge indicates the IR drop is reduced by the vasculature. The whole pulsed charge process is demonstrated in Figure S18. One representative pulse process is shown in Figure 2g for detailed analysis. At the end of the charging process, the lowest voltage of 4-branches vasculature represents the enhanced high-rate charging performance due to the reduced tortuosity and graded average porosity at a certain xy-plane. Such improvement is also consistent with previous study [22].

To provide more physical insights, we use tortuosity and penetration depth to quantify the electrochemical transport capability of different electrode configurations and to analyze the limiting factors of the porous electrodes. Since tortuosity is dependent on porosity, the tortuosity of electrodes is calculated by the local average porosity of each electrode's section as described in Supporting Information Note 4. On the other hand, the penetration depth quantifies how thick the electrode could be effectively used by chemical reaction (Li-ion insertion). This value is derived by solving the partial differential equation groups given by Newmann et al. [19] The final expression is written below:

$$\frac{L}{v} = a_1 \sqrt{\frac{1}{1-\varepsilon}} \qquad (2)$$

where $a_1$ is a constant (see details in Supporting Information Note 5) and $\varepsilon$ is the average porosity of electrode at the certain xy-plane to characterize the local transport properties near the separator-electrode interface. The derivation of this penetration depth is shown in

Supporting Information Note 5. (Here, the geometric factors are the same as the above calculation: *alpha3* = 0.5, *onset3* = 0.14 mm, *b* = 0.005 mm). Specifically, higher tortuosity and/or larger penetration depth lead to high charging capacity. We plot the tortuosity and penetration depth near the separator-electrode interface for different geometry in Figure 2h. Apparently, the tortuosity decreases by introducing channels into the porous electrode, consistent with previous research results. [4-18] The penetration depth increases gradually with more branching channels near the separator-electrode interface by providing a higher average porosity at the xy-plane in this region (eq. 2). The higher average porosity near the separator of vascularized electrode with more branches could be validated by calculating the average porosity in Figure S19. Results from the theoretical models agree well with the electrochemical analysis in Figures 2c-g, which not only proves our bio-inspired idea of vasculature but serves as a practical design guideline for future development.

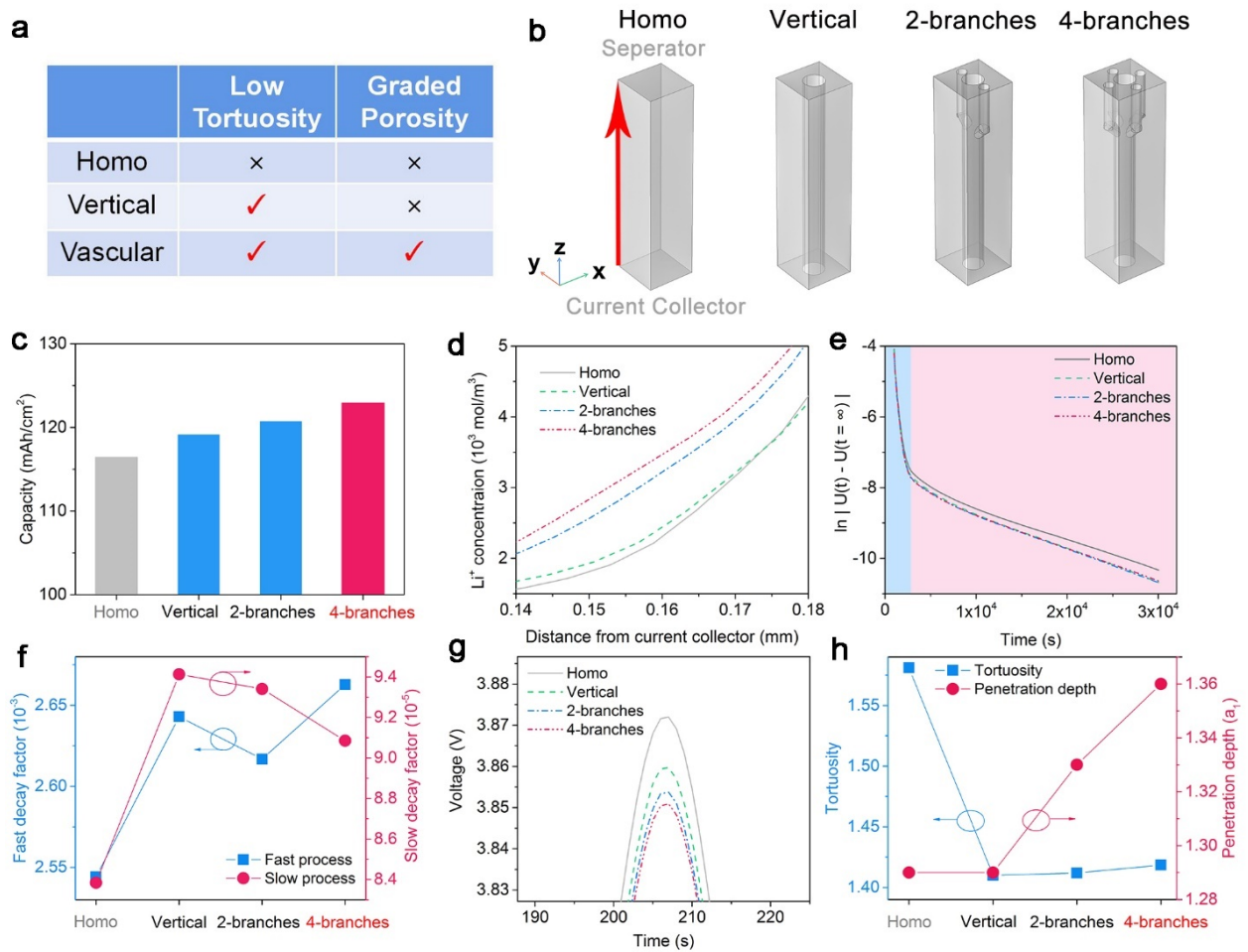

**Figure 2. Electrochemical comparative analysis of different electrode structures.** (a) The table comparing homogenous (traditional) electrodes, vertical channel electrodes, and our vascular channel electrodes. Two critical factors are identified: tortuosity and graded porosity. (b) The unit cells in this calculation. The vascular channels are located near the electrode-separator interface. The red cutline is the edge of the unit cell, which is used for the Li-ion insertion concentration calculation. (c) Charge capacity of various electrode structures under 5C. (d) The zoomed-in figure of the Li ion insertion concentration profile in the x-axis range from 0.14 to 0.18 mm under 10C. The cutline is defined in Figure 2b. (e) DC depolarization test. Two linear diffusion processes are present: the fast kinetic process in blue shadow: 1000 ~ 2000 s and the slow kinetic process in red shadow: 2000 ~ 30000 s). (f) The fitted slopes of different electrode geometries for fast (1000 ~ 2000 s) and slow (2000 ~ 30000 s) diffusion processes, which represent the voltage decay factors. (g) Zoomed-in high C-rate pulse charging figure for one pulse process. (h) Calculated tortuosity and penetration depth with the physical model near the electrode-separator interface concerning different geometries.

**Artificial neural network development and performance**

The success of these specific vasculatures invites a comprehensive optimization among all geometry parameters. More importantly, we desire to obtain the capability to inverse-design the ideal vasculature based on particular practical requirements or limitations. Considering the enormously complex parameter space to describe the vasculature, we utilize deep learning as part of the optimization and design tools. The details of our ANNs structure and the bagging ensemble algorithm are described in Supporting Information Note 6. In Figure 3a, the fully overlapped charging curves (The blue solid lines represent FEM simulation, and the red dash lines represent ANN prediction) under different charging rates and geometries show excellent prediction accuracy of our neural networks. The low mean square error (MSE) loss of ~$10^{-4}$ is another strong evidence of the great prediction (Figure 3b). In Figure 3c, the simulated and predicted capacity values aligned closely with the y = x line with a high $R^2$ value of 0.9995, also suggesting high prediction accuracy. The key advantage of our ANNs is achieving high-fidelity prediction of charging curves with a much smaller demand for computational power compared to the exhaustive combinatorial FEM. As shown in Figure 3d, our deep learning model is 84 times faster than total FEM in screening the parameters and building the total library. Detailed time calculation is shown in Supporting Information Note 6.

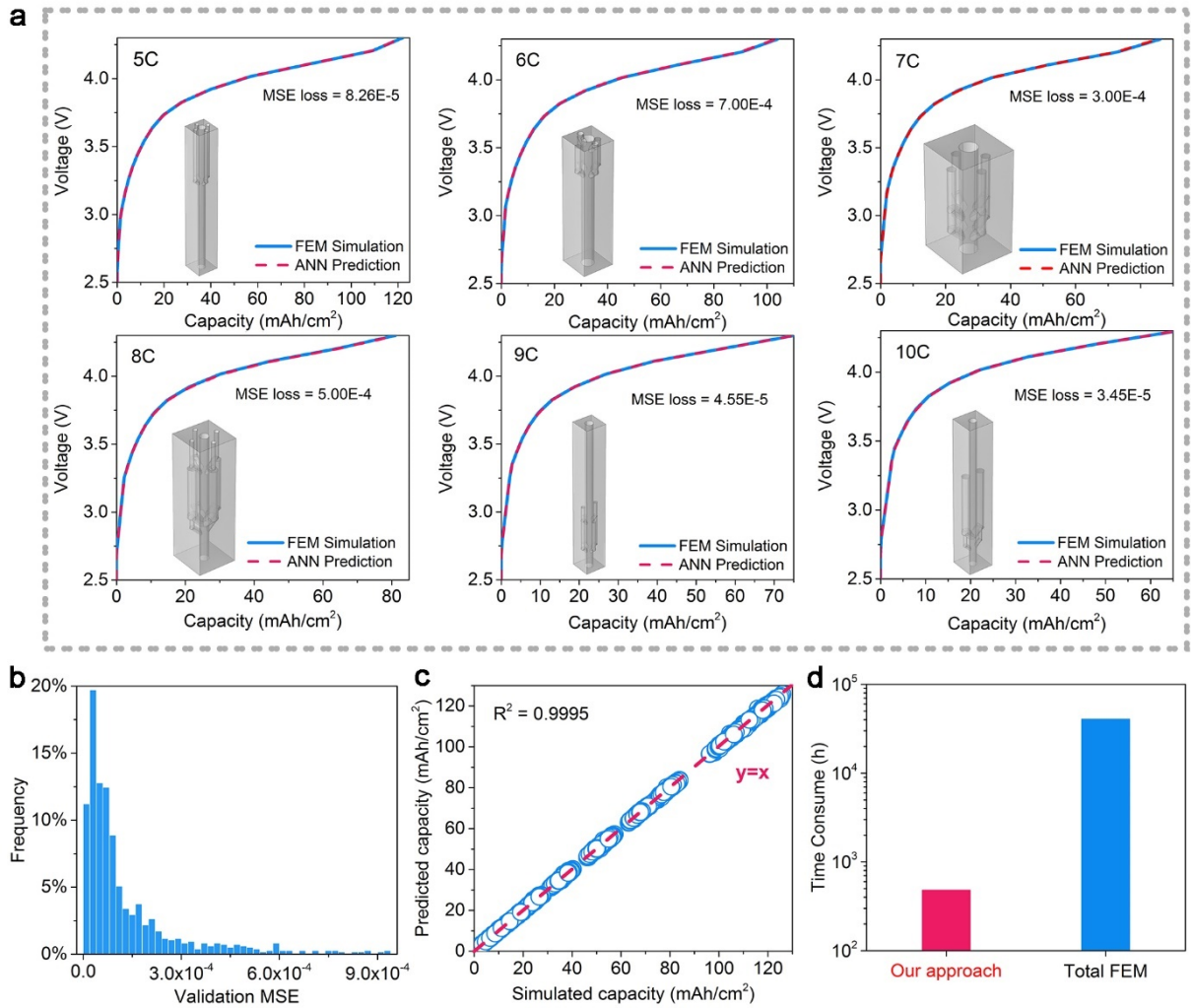

**Figure 3. Artificial neural network training performance.** (a) Comparison of predicted and simulated charging curves under different C-rates and vasculatures. The geometric parameters for these 6 geometries are listed in the table S6. Each curve consists of 20 data points evenly spaced along the y-axis (Figure S10). The mean square error (MSE) loss of the simulation is presented in the upper right corner. (b) The histogram of the validation test MSE. Most of the validation MSEs are smaller than $3.0 \times 10^{-4}$. (c) Correlation between simulated data and ANN predicted data. All data points are aligned near y = x representing the high accuracy of the trained neural network. (d) Computation time for building the total library using our deep learning prediction versus traditional finite element computation.

**Inverse design**

Compared with the forward prediction of the charging curve for each geometry and rate, battery manufacturers are often more interested in understanding what vascular channel structure has the best performance under various kinds of constraints. However, such inverse design problems are generally more challenging because of the complicated geometric parameters of vasculature and the nonlinearity of objective functions [49], especially in time-dependent models such as batteries. Therefore, we take advantage of the forward-prediction computation efficiency and accuracy of our ANNs and construct a total library storing the information such as charging curves, electrode channel structures, charging rates, theoretical capacity, and average porosity at a certain xy-plane. From this total library, we can rank the performance under certain practical requirements/limits of manufacturability, cost, or C-rate. To make the library more comprehensive, we also add other useful electrochemical properties such as energy density, average voltage, and power density of each vascular structure. This approach is inspired by the "fast forward dictionary search (FFDS)", which was first demonstrated by Nadell et al. to be a powerful design tool for all-dielectric metasurfaces [50]. For batteries, the major differences besides the physical laws are the time-dependent behaviors and the emphasis on high charging capacity rather than specific peak features.

For the inverse design problems, after combining a given optimization target with an additional industrial fabrication limit (e.g., the smallest channel radius possible to fabricate, the low power density limitation for efficient charging, etc.), we can easily find the optimal vascular structures by searching in the total library constructed by forward prediction. Such inverse design workflow for finding the optimized geometries adapting certain requirements is demonstrated in Figure 4a. Herein, we give two optimization scenarios: A (find the maximum energy density under 5C and the minimum channel radius is 0.01 mm) and B (find the maximum energy density under 10C and the minimum channel radius is 0.005 mm) for demonstration. Details for calculating power density, energy density, and more specific inverse design method's description are written in the Supporting Information Note 6. The optimal top two geometries among the top three branch number configurations are listed in Table S4. We select the top one geometry for each scenario and draw them in Figure 4b. It is obvious the optimized geometries for high charging capacity and energy density still follow the "more branches near the separator" rule. The porosity profiles analysis for different branches number configurations also support the general rule (Supporting Information Note 7 and 8), making our inverse design method more convincing. Apart from the geometric parameters, our inverse design method can also provide the optimal geometries' corresponding charging curves (Figure 4c, 4d) The different optimal results of branches number configurations between scenario A and B demonstrate the importance and necessity of our inverse design method, because the cells with high charging capacity or energy density may also come at the price of the high overpotential (represented by the high charging power for the same C-rate) that is not desired for charging efficiency. Thus, the inverse design can effectively provide the cell's geometry for different industrial applications that have various technical limitations, such as fabrication size limitation, charging power density limitation, and waste heat generation. With the comprehensive library compiled by the efficient ANN, the inverse problems are no longer challenging even with the highly complicated geometric parameters of vasculature.

**Global sensitivity analysis (Sobol's method)**

In practice, there can be fabrication errors in microchannel geometry and solid matrix porosity that influence the prediction accuracy. In other words, although the vasculatures designed by ANNs and total library work well for COMSOL modeling, the realistic deviation might influence the accuracy for real-world batteries. Therefore, we use Sobol's method to analyze the sensitivity of each parameter on charging capacity to determine the robustness and significance of these parameters (Supporting Information Note 9) [51-53]. Another benefit of sensitivity analysis is to help correlate the theory and simulation in real numbers. In Figure 4e, we can see that the Sobol index for *num3* is the largest, which indicates the number of branches on the onset node 3 has the highest impact on charging capacity. This result agrees well with the theory because the ionic current near the separator-electrode interface region is much higher than that near the current collector. Apart from the number of branches in this region, branch radius (expressed by the ratio to central channel radius *alpha1-3*) will also impact the average porosity at certain xy-plane, although not as significantly as the branch number. If the branch radius is larger near the electrode-separator interface, the average porosity at xy-plane would be increased, and the capacity would be enhanced. Central channel radius *b* is another significant factor because it changes the unit cell size when the areal mass loading and electrode thickness are fixed. If the unit cell is larger, the lateral transport distance of Li-ion from the channel to the corner of the unit cell would be longer, slowing down the ion transport. [4]

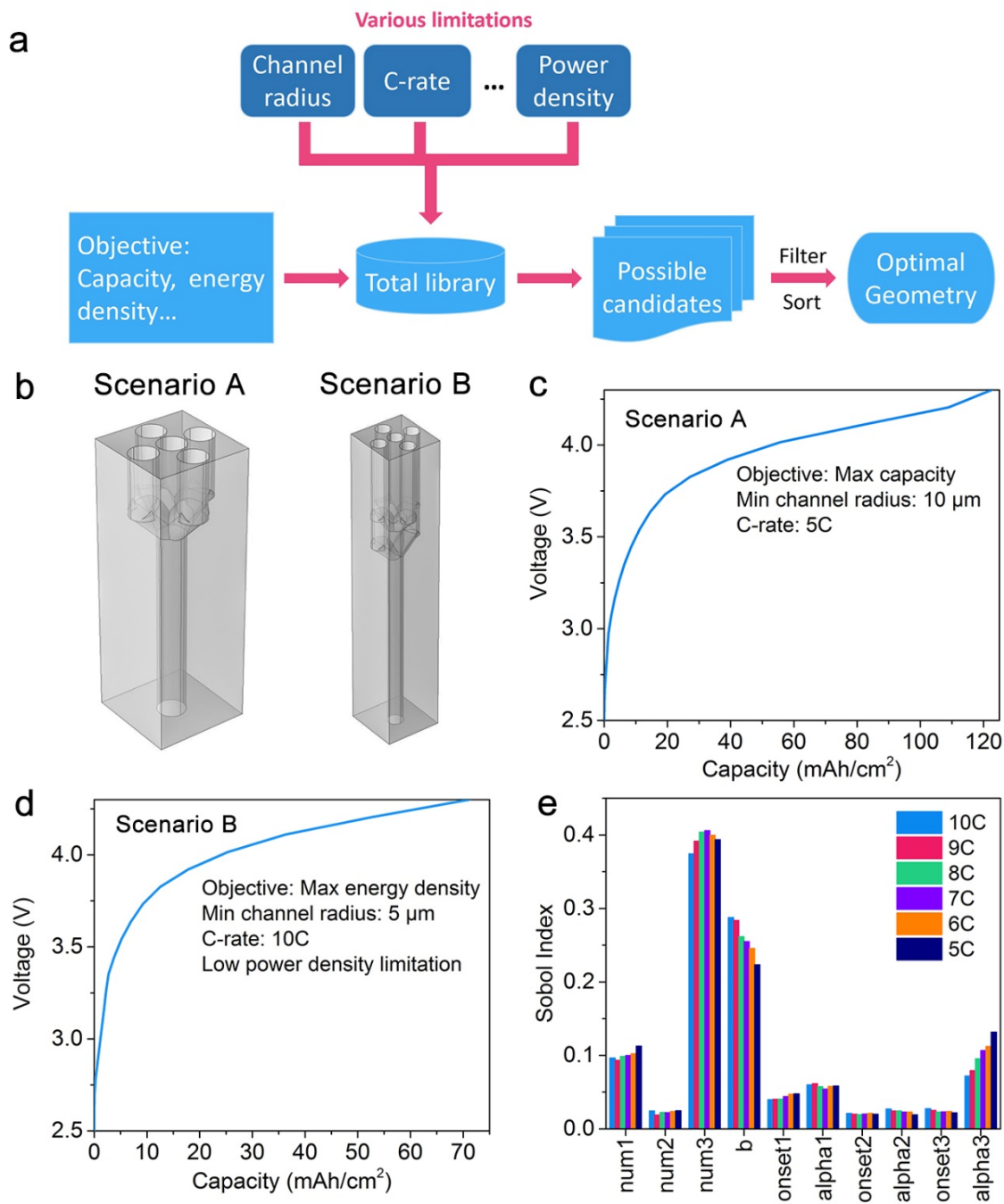

**Figure 4. Customized inverse design package.** (a) Workflow of inverse design. (b) The optimized geometries for scenarios A and B. (c, d) Predicted charging curves for scenarios A and B. (e) Global sensitivity analysis for each geometry parameter shown by Sobol Index.

**Full-cell results**

Based on the enhancement by the vascular channels in graphite anode, it is promising to apply the configurations on both anode and cathode and further boost the fast-charging performance at the full-cell level (Figure 5a). The model computation method is described in Supporting Information Notes 2 and 3. Figure 5b shows the charging capacities for different structured electrodes under 5C and 10C charging rates. The vascular branching channels give the battery a higher improvement than vertical channels, no matter it is anode-structured or double-structured. The best performance is delivered by the double vascular battery, which is as high as 118 mAh/cm$^2$ and 47.6 mAh/cm$^2$ areal capacity under 5C and 10C charging rates, respectively. That means 43.9% and 13.94% improvement compared to homogeneous electrodes are realized (for 5C and 10C charging rates). Even compared with the half vertical channels case, the double vascularized battery still possesses 39.72% and 8.06% improvement under 5C and 10C. Since the overall porosity of all the batteries here is the same, the results again prove the ion transport improvement and IR drop reduction induced by vascular structures. The considerable improvement shines the light on a fast-charging battery in real industrial production by the bio-inspired vasculature idea.

The U.S. Advanced Battery Consortium goal for fast-charging electric vehicles' batteries is to charge to 80% capacity within 15 mins. [1] If we assume a constant charging process, the charging rate should be at least 3.2C. In Figure 5c, such a current is used to charge the homogeneous porous electrodes (40% overall porosity) battery and double vascular full-cell batteries (both cathode and anode are vascularized with 40% overall porosity). The double vascular full-cell battery uses the same vascular geometry as listed in table S3. The capacity of the double vascular full cell is 211.9 mAh/cm$^2$, almost 66% larger than that of the homogeneous electrode (128.1 mAh/cm$^2$). All electrochemical parameters are the same as in table S2.

We further test the vascularized full cells with pulse charging. (Figure 5d) The pulse charging current configuration is set as the smooth periodic piece-wise function to ensure convergence of computation. The cell is charged under 15C charging rate for 0.2 s and then rest for 0.2 s under zero current. The process is repeated until the voltage reaches the cutoff voltage (4.3 V). Because of the low tortuosity and graded porosity, the double vascular full-cell battery exhibits a larger capacity (pulse charging for 22.98 s) than a regular homogeneous counterpart (pulse charging for 16.98 s). Such 1.35-time improvement clearly demonstrate the effectiveness of vasculature design for high-rate charging performance.

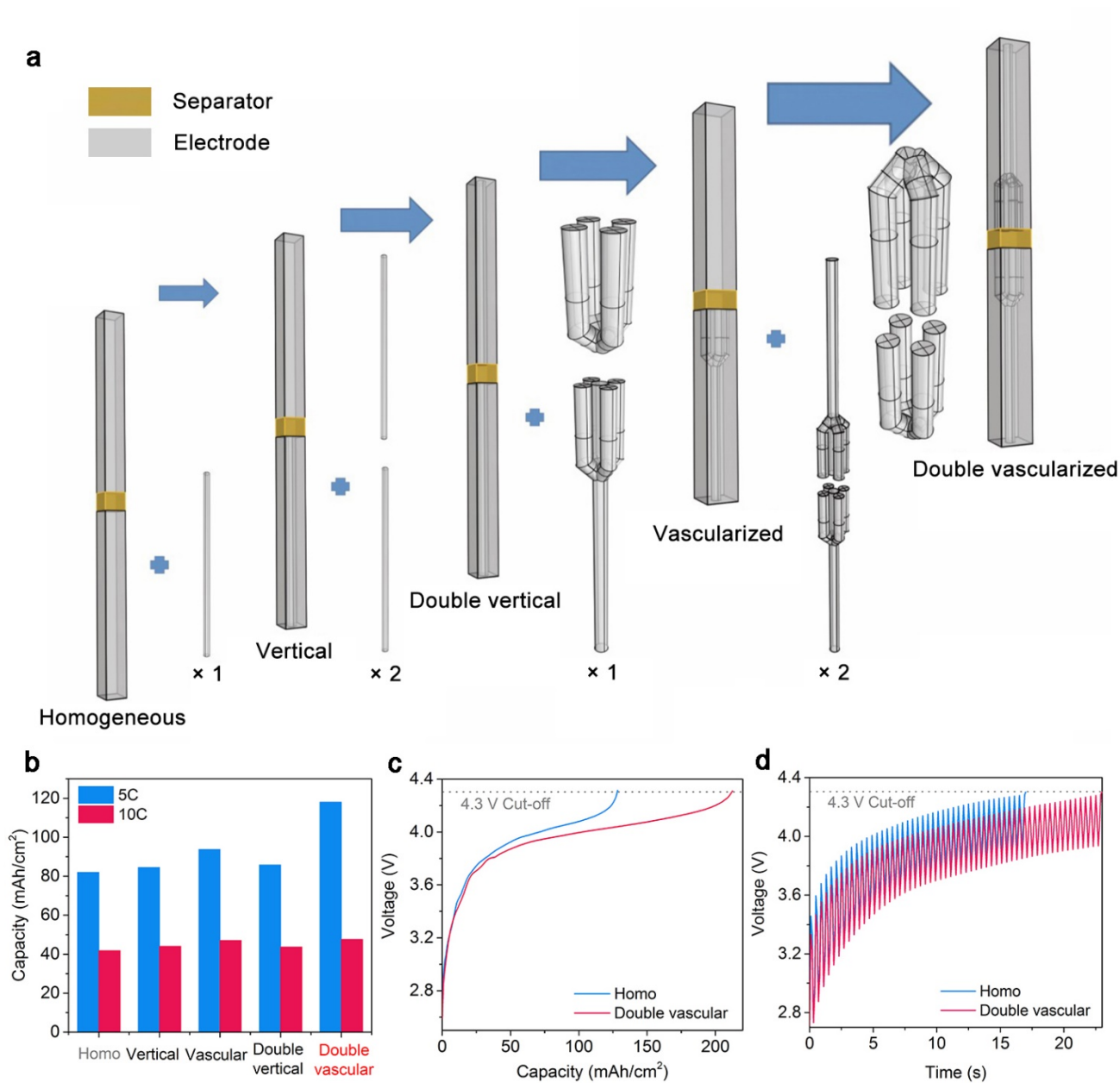

**Figure 5. The full-cells simulation for real application scenarios.** (a) The schematic for designing full cells with traditional vertical channels and vascular channels. Each of the cases represents the unit cell for that kind of geometry. The charging performance enhances gradually by introducing more vasculatures. (b) Charging capacity comparison between different configurations under 5C and 10C charging rates. The vascularized structure shows a significant improvement, and the double vascular structure possesses the best performance. (c) Charging curves comparison under the U.S. Advanced Battery Consortium's goal (3.2C). (d) High-rate pulse charged curves comparison.

## 3. Conclusions

In this article, the bio-inspired vasculature approach to address the performance degradation of Li-ion batteries under a high charging rate is presented. The fundamental electrochemical theory explains that the vascularized porous electrode improves the fast-charging performance in two ways. First, it reduces the tortuosity of the electrode. Second, the hierarchical branches of vascular structures create gradual porosity distribution profiles. Among them, the graded profiles with high average porosity near the separator-electrode interface will improve the performance by suppressing the IR-drop of the whole system. We demonstrated an efficient deep-learning prototype to predict and inver-design the vascular structures under different criteria. Finally, full-cell analysis was conducted and showed vasculature can realize a 66% improvement under 3.2C constant current charge and a 35% improvement of charging time under 15C pulse-charge, compared to the homogeneous batteries. Looking beyond batteries, this work could also potentially serve as an innovative methodology for optimizing the topological design of complex 3-dimensional transportation systems which cannot be solved analytically, such as flow-based gas catalysis [54], biological diffusion systems [55], and other reactive flow problems.

## 4. Method

**Finite element computation**

The battery charging simulation is conducted using COMSOL Li-ion battery Module. All electrochemical simulation parameters and geometric model are described in the Supporting Information Note 1-3. The massive parametric sweeping is achieved by Cluster Sweep feature and the software is run on 400 CPUs supported by Duke Compute Cluster.

**Deep learning**

The artificial neural networks are constructed using PyTorch (1.9.0 + CUDA 10.2 version) open-source machine learning library. All details about the structure, input parameters and output results are described in the Supporting Information 6. The neural network training is conducted on the GPU supported by Duke Compute Cluster.

**Acknowledgment**

The authors want to thank Prof. Willie Padilla and Dr. Omar Khatib for the valuable discussion regarding artificial neural network development and inverse design strategy. We also thank Prof. Jon Herman (UC Davis, Department of Civil and Environmental Engineering) for kindly helping with the sensitivity analysis and COMSOL Inc. for the technical support on simulation. Hao Zhang (Duke University, Civil and Environmental Engineering Department) also provided a useful discussion on the data analysis. This work is supported by North Carolina Space Grant. The authors also thank Pratt School of Engineering at Duke University for the funding support.

**Author Contributions:**

P.C.H. and C.S. conceived the idea. C.S. performed the finite element simulation and electrochemical theoretical analysis. C.S. and Y.Y.L. performed the neural network training and prediction. C.S. and Y.Y.L. performed inverse design optimization. C.S. performed the global sensitivity analysis. G.H. helped with the design of vascular structures. K.W. helped with porosity distribution analysis. X.L. helped with data processing and presenting. W.X. helped with figures. C.S., Y.Y.L., and P.C.H. wrote the manuscript with input from all co-authors.

**Competing interests:**

The authors declare no competing financial interests.

**Data and materials availability:**

All data needed to support the conclusions in the paper are present in the manuscript and/or the Supplementary Materials. Additional data related to this paper may be requested from the corresponding author upon request.

# References

[1] Liu, Y., Zhu, Y., and Cui, Y. (2019). Challenges and opportunities towards fast-charging battery materials. Nature Energy 4, 540-550.

[2] Rogge, M., Wollny, S., and Sauer, D. (2015). Fast Charging Battery Buses for the Electrification of Urban Public Transport—A Feasibility Study Focusing on Charging Infrastructure and Energy Storage Requirements. Energies 8, 4587-4606.

[3] Christopher J., M., Shabbir, A., Ira, B., Andrew, B., Barney, C., Fernando, D., Eric J., D., Andrew N., J., Matthew, K., David, H., et al. (2017). Enabling Fast Charging: A Technology Gap Assessment. No INL/EXT-17-41638 Idaho National Lab(INL), Idaho Falls, ID (United States).

[4] Bae, C.J., Erdonmez, C.K., Halloran, J.W., and Chiang, Y.M. (2013). Design of battery electrodes with dual-scale porosity to minimize tortuosity and maximize performance. Adv. Mater. 25, 1254-1258.

[5] Li, L., Erb, R.M., Wang, J., Wang, J., and Chiang, Y.-M. (2019). Fabrication of Low-Tortuosity Ultrahigh-Area-Capacity Battery Electrodes through Magnetic Alignment of Emulsion-Based Slurries. Advanced Energy Materials 9, 1802472.

[6] Amin, R., Delattre, B., Tomsia, A.P., and Chiang, Y.-M. (2018). Electrochemical Characterization of High Energy Density Graphite Electrodes Made by Freeze-Casting. ACS Applied Energy Materials 1, 4976-4981.

[7] Delattre, B., Amin, R., Sander, J., De Coninck, J., Tomsia, A.P., and Chiang, Y.-M. (2018). Impact of Pore Tortuosity on Electrode Kinetics in Lithium Battery Electrodes: Study in Directionally Freeze-Cast LiNi0.8Co0.15Al0.05O2(NCA). Journal of The Electrochemical Society 165, A388-A395.

[8] Sander, J.S., Erb, R.M., Li, L., Gurijala, A., and Chiang, Y.M. (2016). High-performance battery electrodes via magnetic templating. Nature Energy 1, 1-7.

[9] Billaud, J., Bouville, F., Magrini, T., Villevieille, C., and Studart, A.R. (2016). Magnetically aligned graphite electrodes for high-rate performance Li-ion batteries. Nature Energy 1, 1-6.

[10] Huang, C., and Grant, P.S. (2018). Coral-like directional porosity lithium ion battery cathodes by ice templating. Journal of Materials Chemistry A 6, 14689-14699.

[11] Zhang, Y., Luo, W., Wang, C., Li, Y., Chen, C., Song, J., Dai, J., Hitz, E.M., Xu, S., Yang,

C., et al. (2017). High-capacity, low-tortuosity, and channel-guided lithium metal anode. Proc Natl Acad Sci U S A 114, 3584-3589.

[12] Shen, F., Luo, W., Dai, J., Yao, Y., Zhu, M., Hitz, E., Tang, Y., Chen, Y., Sprenkle, V.L., Li, X., and Hu, L. (2016). Ultra-Thick, Low-Tortuosity, and Mesoporous Wood Carbon Anode for High-Performance Sodium-Ion Batteries. Advanced Energy Materials 6.

[13] Chen, C., Zhang, Y., Li, Y., Kuang, Y., Song, J., Luo, W., Wang, Y., Yao, Y., Pastel, G., Xie, J., and Hu, L. (2017). Highly Conductive, Lightweight, Low‐Tortuosity Carbon Frameworks as Ultrathick 3D Current Collectors. Advanced Energy Materials 7.

[14] Li, Y., Fu, K.K., Chen, C., Luo, W., Gao, T., Xu, S., Dai, J., Pastel, G., Wang, Y., Liu, B., et al. (2017). Enabling High-Areal-Capacity Lithium-Sulfur Batteries: Designing Anisotropic and Low-Tortuosity Porous Architectures. ACS Nano 11, 4801-4807.

[15] Zhao, Z., Sun, M., Chen, W., Liu, Y., Zhang, L., Dongfang, N., Ruan, Y., Zhang, J., Wang, P., Dong, L., et al. (2019). Sandwich, Vertical‐Channeled Thick Electrodes with High Rate and Cycle Performance. Advanced Functional Materials 29.

[16] Lu, L.L., Lu, Y.Y., Xiao, Z.J., Zhang, T.W., Zhou, F., Ma, T., Ni, Y., Yao, H.B., Yu, S.H., and Cui, Y. (2018). Wood-Inspired High-Performance Ultrathick Bulk Battery Electrodes. Adv Mater 30, e1706745.

[17] Luo, J., Yao, X., Yang, L., Han, Y., Chen, L., Geng, X., Vattipalli, V., Dong, Q., Fan, W., Wang, D., and Zhu, H. (2017). Free-standing porous carbon electrodes derived from wood for high-performance Li-O2 battery applications. Nano Research 10, 4318-4326.

[18] Kuang, Y., Chen, C., Kirsch, D., and Hu, L. (2019). Thick Electrode Batteries: Principles, Opportunities, and Challenges. Advanced Energy Materials 9.

[19] Newman, J., and Thomas-Alyea , K.E. (2012) Electrochemical systems. (John Wiley & Sons).

[20] Chiang, Y.M., and Hellweg, B. (2001). Reticulated and controlled porosity battery structures. US Patent, No. 7,553,584. 30 Jun. 2009.

[21] Du, Z., Wood, D.L., Daniel, C., Kalnaus, S., and Li, J. (2017). Understanding limiting factors in thick electrode performance as applied to high energy density Li-ion batteries. Journal of Applied Electrochemistry 47, 405-415.

[22] Ramadesigan, V., Methekar, R.N., Latinwo, F., Braatz, R.D., and Subramanian, V.R.

(2010). Optimal Porosity Distribution for Minimized Ohmic Drop across a Porous Electrode. Journal of The Electrochemical Society 157, A1328.

[23] White, M.C., Marin, D.B., Brazeal, D.V., and Friedman, W.H. (1997). The evolution of organizations: Suggestions from complexity theory about the interplay between natural selection and adaptation. Human Relations 50, 1383–1401.

[24] West, G.B., Brown, J.H., and Enquist, B.J. (1999). The Fourth Dimension of Life: Fractal Geometry and Allometric Scaling of Organisms. Science 284, 1677-1679.

[25] Murray, C.D. (1926). The physiological principle of minimum work: I. The vascular system and the cost of blood volume. Proceedings of the National Academy of Sciences of the United States of America 12, 207–214.

[26] Sherman, T.F. (1981). The Meaning of Murray's Law. J. Gen. Physiol. 78, 431-453.

[27] Hetz, S.K., and Bradley, T.J. (2005). Insects breathe discontinuously to avoid oxygen toxicity. Nature 433, 516-519.

[28] Zheng, X., Shen, G., Wang, C., Li, Y., Dunphy, D., Hasan, T., Brinker, C.J., and Su, B.L. (2017). Bio-inspired Murray materials for mass transfer and activity. Nat Commun 8, 14921.

[29] Guo, N., Leu, M.C., and Koylu, U.O. (2014). Bio-inspired flow field designs for polymer electrolyte membrane fuel cells. International Journal of Hydrogen Energy 39, 21185-21195.

[30] Huang, J.-H., Kim, J., Agrawal, N., Sudarsan, A.P., Maxim, J.E., Jayaraman, A., and Ugaz, V.M. (2009). Rapid fabrication of bio-inspired 3D microfluidic vascular networks. Advanced Materials 21, 3567-3571.

[31] Kayla, M. 3D printed bioinspired vascularized polymers (The University of Maine).

[32] Marquis, K., Chasse, B., Regan, D.P., Boutiette, A.L., Khalil, A., and Howell, C. (2020). Vascularized Polymers Spatially Control Bacterial Cells on Surfaces. Adv Biosyst 4, e1900216

[33] Howell, C., Vu, T.L., Lin, J.J., Kolle, S., Juthani, N., Watson, E., Weaver, J.C., Alvarenga, J., and Aizenberg, J. (2014). Self-replenishing vascularized fouling-release surfaces. ACS Appl Mater Interfaces 6, 13299-13307.

[34] Hatton, B.D., Wheeldon, I., Hancock, M.J., Kolle, M., Aizenberg, J., and Ingber, D.E. (2013). An artificial vasculature for adaptive thermal control of windows. Solar Energy Materials and Solar Cells 117, 429-436.

[35] Zhu, X., Zhao, C., Wang, X., Zhou, Y., Hu, P., and Ma, Z.-D. (2019). Temperature-

constrained topology optimization of thermo-mechanical coupled problems. Engineering Optimization 51, 1687-1709.

[36] Xue, R., Li, R., Du, Z., Zhang, W., Zhu, Y., Sun, Z., and Guo, X. (2017). Kirigami pattern design of mechanically driven formation of complex 3D structures through topology optimization. Extreme Mechanics Letters 15, 139-144.

[37] Kim, C., and Sun, H. (2012). Topology optimization of gas flow channel routes in an automotive fuel cell. International Journal of Automotive Technology 13, 783-789.

[38] Song, X., Diaz, A.R., and Benard, A. (2013). A 3D topology optimization model of the cathode air supply channel in planar solid oxide fuel cell. 10th World Congress on Structural and Multidisciplinary Optimization.

[39] Xu, T. Topology optimization of lithium ion batteries: How to maximize the discharge capacity by changing the geometry (Delft University of Technology).

[40] Bottou, L., Curtis, F.E., and Nocedal, J. (2018). Optimization Methods for Large-Scale Machine Learning. SIAM Review 60, 223-311.

[41] Zhang, Y., Tang, Q., Zhang, Y., Wang, J., Stimming, U., and Lee, A.A. (2020). Identifying degradation patterns of lithium ion batteries from impedance spectroscopy using machine learning. Nat Commun 11, 1706.

[42] Attia, P.M., Grover, A., Jin, N., Severson, K.A., Markov, T.M., Liao, Y.H., Chen, M.H., Cheong, B., Perkins, N., Yang, Z., et al. (2020). Closed-loop optimization of fast-charging protocols for batteries with machine learning. Nature 578, 397-402.

[43] Bhowmik, A., and Vegge, T. (2020). AI Fast Track to Battery Fast Charge. Joule 4, 717-719.

[44] Finegan, D.P., Zhu, J., Feng, X., Keyser, M., Ulmefors, M., Li, W., Bazant, M.Z., and Cooper, S.J. (2021). The Application of Data-Driven Methods and Physics-Based Learning for Improving Battery Safety. Joule 5, 316-329.

[45] Tian, J., Xiong, R., Shen, W., Lu, J., and Yang, X.-G. (2021). Deep neural network battery charging curve prediction using 30 points collected in 10 min. Joule 5, 1521-1534.

[46] Li, W., Zhu, J., Xia, Y., Gorji, M.B., and Wierzbicki, T. (2019). Data-Driven Safety Envelope of Lithium-Ion Batteries for Electric Vehicles. Joule 3, 2703-2715.

[47] Kandler, S. and Wang, C.Y. (2006) Power and thermal characterization of a lithium-ion

battery pack for hybrid-electric vehicles. Journal of power sources 160, 662-673.

[48] Rui, Z., Liu, L. and Ma, F. (2020) Cathode Chemistries and Electrode Parameters Affecting the Fast Charging Performance of Li-Ion Batteries. Journal of Electrochemical Energy Conversion and Storage 17.

[49] Chavent, G. (2010). Nonlinear least squares for inverse problems: theoretical foundations and step-by-step guide for applications. (Springer Science & Business Media).

[50] Nadell, C.C., Huang, B., Malof, J.M., and Padilla, W.J. (2019). Deep learning for accelerated all-dielectric metasurface design. Opt Express 27, 27523-27535.

[51] Sobol, I.M. (2001). Global sensitivity indices for nonlinear mathematical models and their Monte Carlo estimates. Math. and Comput. Simul. 55, 271–280.

[52] Herman, J., and Usher, W. (2017). SALib: An open-source Python library for Sensitivity Analysis. The Journal of Open Source Software 2, 97.

[53] Do, N.C., and Razavi, S. (2020). Correlation Effects? A Major but Often Neglected Component in Sensitivity and Uncertainty Analysis. Water Resources Research 56, e2019WR025436.

[54] Wheeler, D.G., Mowbray, B.A.W., Reyes, A., Habibzadeh, F., He, J., and Berlinguette, C.P. (2020). Quantification of water transport in a CO2 electrolyzer. Energy & Environmental Science 13, 5126-5134.

[55] Tomasina, C., Bodet, T., Mota, C., Moroni, L., and Camarero-Espinosa, S. (2019). Bioprinting Vasculature: Materials, Cells and Emergent Techniques. Materials (Basel) 12, 2701.

# Supporting Information

**Note 1. Geometry Model description**

The porous graphite anode model is built to simulate the different performances of geometric structures. Here, we investigate the influence of vascular channels. Thus, only the geometric structure of channels is changed in different models. The thickness of the anode *h* is fixed as 200 μm. We also control the overall density $D_{overall}$ to equal 60% for all anodes with different channel structures. Note that density describes the volume fraction of solid matrix in the whole porous electrode area and is equal to '100% − porosity'. Since the configuration of the branched channels would be periodically arranged along the cross-sectional area of the electrode, we only simulate one unit cell to represent the performance of the whole periodic structure. Fixing the overall density or overall porosity can guarantee that any change in performance would only be caused by the variance of the structure of the branched channels. The vascular arrangement is shown in Figure S1. Figure S1a is the cross-section of an anode with vascular channels from the side view. Figure S1b is the top view of the anode, which is also the interface between the anode and the separator. Figure S1c in the bottom view of the anode, which is the interface between anode and electric ground current collector. For a homogeneous porous electrode, the density of the electrode solid-phase matrix $D_{electrode}$ is set to be 60%. For the electrode with a branched structure, the porosity of channels, or void space, is equal to 100%. To maintain the overall density of the whole unit cell equal to 60%, we must increase the density of the electrode solid-phase matrix $D_{electrode}$ to 70%. With the assumptions above, we solve the parameters' dependence in Figure S1. For the branches in Figure S1, we have one vertical central channel starting from the top to bottom with radius *b*. Conventional vascular structures need to branch out of the central channel. We set three nodes for branches to grow out. The distances of each node from the top of the anode are described as *onset1*, *onset2*, and *onset3*. On each node, we have a different number of branches (0, 2, or 4 branches) and they are written as *num1*, *num2*, and *num3*. The branch starts to grow from its node along with 45-degrees concerning the central channel and stretches out to the position of 70% of half of the unit cell length. All the branches reaching out from the same node have the same radius. These radiuses are described as $r_1$, $r_2$, and $r_3$. The branch radiuses are set to be a half or quarter of the central channel radius *b*. The unit cell size is *a*. To find the relationship between these 11 parameters, we need to combine them into one equation which represents the overall density, or overall porosity. To do that, we have the derivation below. The volume of the space inside the channel could be calculated as a function of *a*, *b*, $r_1$, $r_2$, $r_3$, and *h*.

$$V_{channel} = f(a, b, r_{1-3}, onset_{1-3}, num_{1-3}, h) \quad (1)$$

The volume of the unit cell is calculated by

$$V_{unit} = a^2 h \quad (2)$$

The volume of the electrode solid phase matrix is expressed as

$$V_{electrode} = V_{unit} - V_{channel} \quad (3)$$

With the volume of each component of the electrode, we could write the overall porosity as

$$D_{overall} = \frac{D_{electrode} \cdot V_{electrode}}{V_{unit}} = 0.6 \quad (4)$$

By solving the equation (4) explicitly for *a*, we could get the relationship between *a* and the other 10 geometry parameters. Above all is the overall porosity constrain for 11 geometric parameters. Therefore, if we vary the geometry factors in the rational range (for the geometry structure to be successfully formed), the overall porosity of the vascular electrode could be fixed to 40%.

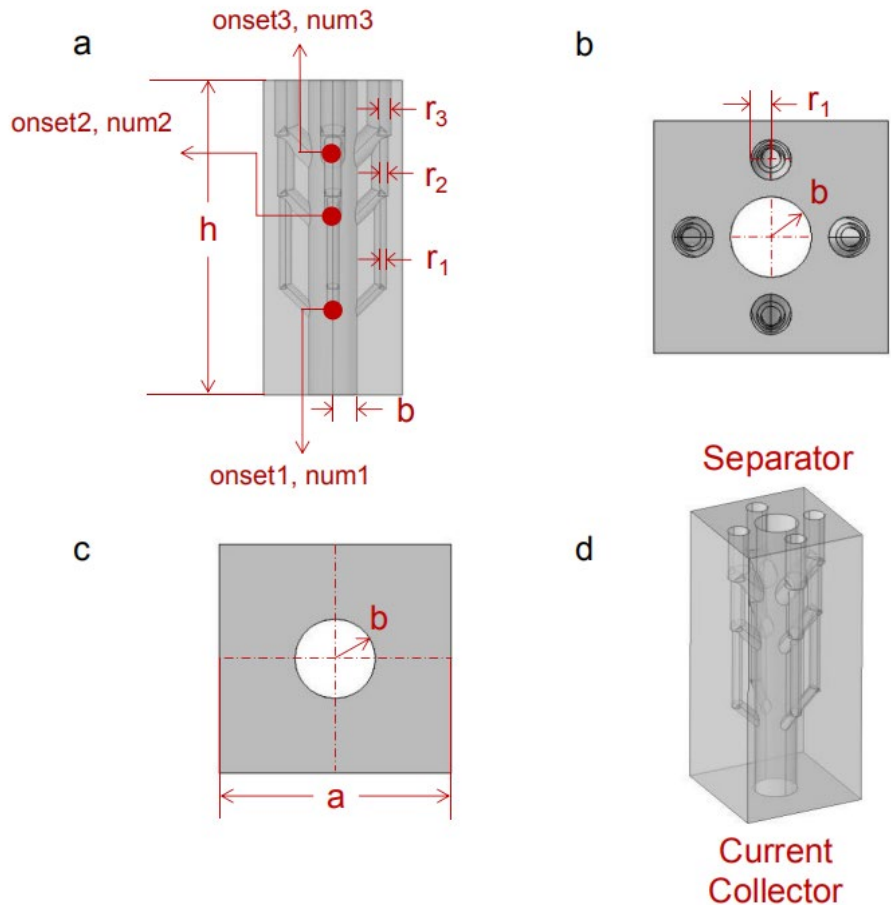

Figure S1. Geometry schematic for an anode unit cell.

**Note 2. Electrochemistry battery performance modeling**

The finite element battery model is built to simulate the battery performance with vascular channels. All the computations are conducted by the COMSOL Li-ion battery module. On the top of the anode in Figure S1d, there would be a thin layer of $LiPF_6$ in 3:7 EC:EMC electrolyte. A $LiCoO_2$ (LCO) cathode is put on the top of the electrolyte layer. In contact, these three layers form the electrochemical cell. According to previous literature, exchange current density is equal to 36 A/m$^2$ and 26 A/m$^2$ for graphite and LCO, respectively. We set the exchange current density value for graphite to be 36 A/m$^2$. [3] On the other hand, the focus of our computation is to test the performance of the anode, so we eliminate the influence of cathode by setting the exchange current for LCO to be 260 A/m$^2$, ten times higher than the normal value (26 A/m$^2$). In this way, the limitation of the cathode could be ignored. Thus, the cell is regarded as a pseudo-half-cell to study anode performance. All electrochemistry simulations follow the partial differential equations group for battery systems developed by John Newman. [1] Other materials parameters all come from the COMSOL materials library, which is shown in Table S1. The training dataset for the neural network is simulated by the pseudo-half-cell model mentioned above. The overall porosity for a positive homogeneous electrode is 30% (for a sufficient amount of LCO), and the overall porosity for a negative electrode solid phase matrix is also 30%.

For full-cell simulation, (Figure 5 Double vascular) all parameters and boundary conditions are set to simulate the realistic full-cell Li-ion battery working performance. The exchange current density is obtained from previous literature. (26 A/m$^2$ for LCO electrode and 36 A/m$^2$ for graphite electrode) [3] For the onset position of vascular structures, we set the onset3 equal to 0.14 mm. To simulate the experimental limits, we choose the channel radius as 5 μm, which is close to the current fabrication limitation of 6 μm from the previous experiment performed by Bae, Chang‐Jun, et al. [4] As for electrode density, we set the overall density for positive and negative electrodes equal to 60%. For a homogeneous electrode, the solid phase matrix density is 60%. For electrodes with vascular channels, the density of the solid phase matrix is set to be 70% to fix the overall density to be 60%. Note that all electrode fabrication must contain polymer binders and conductive fillers as additives. Here, we set the volume fraction of additives equal to 0.1. Therefore, the summation of the electrode of volume fraction and electrolyte volume fraction is equal to 0.9 in all electrodes. All electrochemical properties and boundary conditions are listed in table S2. Also, the geometry parameters are in Table S3. Interpolate value plots are shown in Figure S2~S5. To simulate the charging process, we set the initial cell voltage to be 2.5 V and cut-off voltage to be 4.3 V. The computation will stop once the voltage hits the cut-off condition. For massive training dataset computation, the Cluster Sweep function is used to connect COMSOL to Duke Compute Cluster with 400 cores of CPUs.

| Properties (LCO) | Value |
|---|---|
| Electrical conductivity | 10 S/m [2] |
| Diffusion coefficient | 5e-13 m²/s |
| Reference concentration | 56250 mol/m³ |
| Density | 5000 kg/m³ |
| Minimum State-of-Charge | 0.43 |
| Maximum State-of-Charge | 1 |
| Properties (Graphite) | Value |
| Electrical conductivity | 100 S/m |
| Diffusion coefficient | $1.4523 \times 10^{-13} \cdot \exp(68025.7/8.314 \cdot (1/(T_{ref}/1[K]) - 1/(T_2/1[K])))$ m²/s |
| Reference concentration | 31507 mol/m³ |
| Density | 2300 kg/m³ |
| Minimum State-of-Charge | 0 |
| Maximum State-of-Charge | 0.98 |
| Properties (LiPF$_6$ in 3:7 EC:EMC) | Value |
| Diffusion coefficient | $D_{int}(c/1[mol/m^3]) \cdot \exp(16500/8.314 \cdot (1/(T_{ref}/1[K]) - 1/(T_2/1[K])))$ |
| Electrolyte Conductivity | $\sigma_{int}(c/1[mol/m^3]) \cdot \exp(4000/8.314 \cdot (1/(T_{ref}/1[K]) - 1/(T_2/1[K])))$ |
| Transport number | $t_{int}(c/1[mol/m^3])$ |
| Activity dependence | $a_{int}(c/1[mol/m^3]) \cdot \exp(-1000/8.314 \cdot (1/(T_{ref}/1[K]) - 1/(T_2/1[K])))$ |
| Variables | Value |
| $T_{ref}$ | 298 K |
| $T_2$ | min(393.15,max(T,223.15)) K |
| T | 293.15 K |
| c | Electrolyte salt concentration |

#$D_{int}$, $\sigma_{int}$, $t_{int}$ and $a_{int}$ are interpolate values as the function of T. (Figure S2 ~ S5)

Table S1. Materials properties from COMSOL library.

| Variables | Value |
|---|---|
| Particle size | 1e-6 m [2] |
| LCO exchange current density | 260 A/m$^2$ # (26 A/m$^2$ for full-cell [3]) |
| Graphite exchange current density | 36 A/m$^2$ [3] |
| Initial electrolyte salt concentration | 1000 mol/m$^3$ # |
| Electrode volume fraction (half-cell) | vascular electrode: 0.7#, homogeneous electrode: 0.6# |
| Additives volume fraction (half-cell) | 0.05# |
| Electrolyte volume fraction (half-cell) | vascular electrode: 0.25#, homogeneous electrode: 0.35# |
| Electrode volume fraction (full-cell) | vascular electrode: 0.7#, homogeneous electrode: 0.6# |
| Additives volume fraction (full-cell) | 0.1# |
| Electrolyte volume fraction (full-cell) | vascular electrode: 0.2#, homogeneous electrode: 0.3# |
| Anodic transfer coefficient | 0.5# |
| Cathodic transfer coefficient | 0.5# |
| Electrolyte reference concentration | 1000 mol/m$^3$ # |
| Separator Porosity | 0.5 for full-cell [2] (no separator for half-cell) |

#Assumed data

Table S2. Electrochemical properties.

| Parameter | Value |
|---|---|
| Separator thickness | 0.02 mm |
| Electrode thickness (cathode and anode for all model) | 0.2 mm |
| Channel radius (vertical model) | 0.005 mm |
| Unit cell size (vertical model) | 0.023447 mm |
| Central channel radius (004 vascular model) | 0.005 mm |
| $\alpha_3$ (004 vascular model) | 1 (This means branch radius is equal to 0.005 mm) |
| $onset_3$ (004 vascular model) | 0.14 mm (Distance from the current collector) |
| Unit cell size (004 vascular model) | 0.034448 mm |

Table S3. Geometry parameters for full-cell simulation

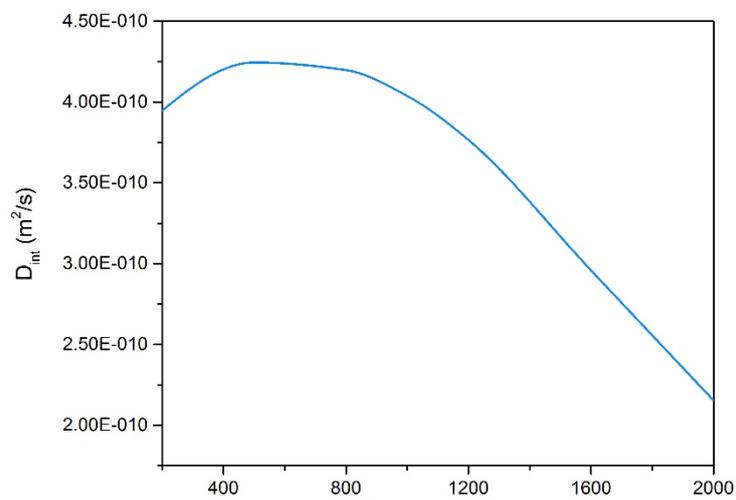

Figure S2 Interpolate value plot of $D_{int}$. (From COMSOL materials library)

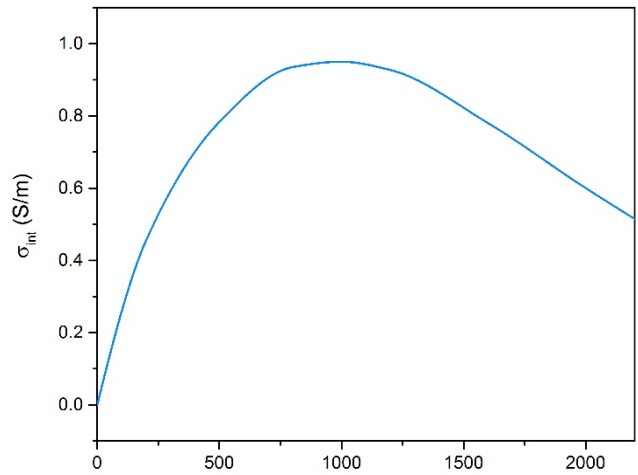

Figure S3. Interpolate value plot of $\sigma_{int}$. (From COMSOL materials library)

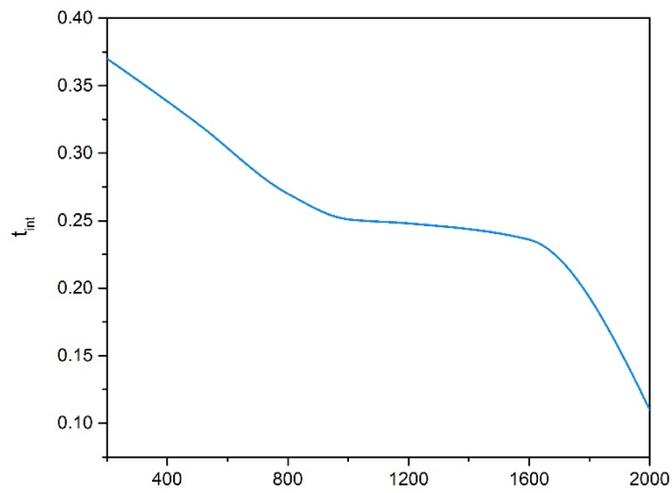

Figure S4. Interpolate value plot of $t_{int}$. (From COMSOL materials library)

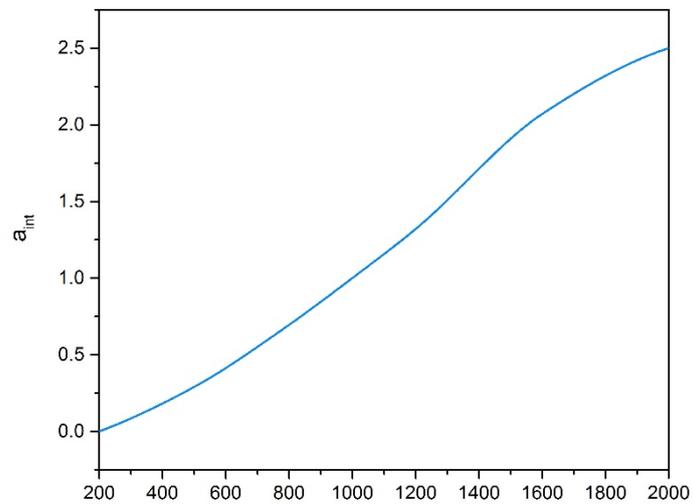

Figure S5. Interpolate value plot of $a_{int}$. (From COMSOL materials library)

**Note 3. 1C Current definition**

To study the high charging rate problem, it is necessary to define the 1C current. First, we need to define the cell theoretical areal capacity when the battery is fully charged.

$$Q = c_{ref} \cdot F \cdot \rho \cdot (SOC_{max} - SOC_{min}) \cdot h$$

Where $c_{ref}$ is the reference concentration of Li-ion in the electrode, F is Faraday constant, $\rho$ is the electrode density, h is the thickness of the electrode. All these variables are defined with respect to the graphite anode. The density of the graphite electrode is always set to be 0.6 for homogeneous and vascular electrodes.

Since the capacity defined above is the areal capacity, we can directly get the 1C current density by dividing areal capacity by 1 hour charging time as below:

$$i_{1C} = \frac{Q}{3600\,[s]}$$

The $c_{ref}$ of graphite electrode is 31507 mol/m³, and the $SOC_{max}$ and $SOC_{min}$ of it are 0.98 and 0. The thickness of the electrode is 200 μm. Since we fix the overall porosity of the anode in the training dataset half-cell model to be 0.4, the electrode density is 0.6. Thus, the current density of 1C is calculated as 99.305 A/m^2. However, the above definition is for half-cell simulation where vascular channels are located in the graphite electrode. The input current density is defined on the LCO electrode, so the current collector is just the bottom area of the cubic electrode with $a^2$ area value. (Figure S6) For full-cell simulation, if the vascular channels are defined in the LCO cathode, the central channel will penetrate through the whole LCO electrode. The bottom area of the LCO electrode (which is also the current collector for electric current density) would equal to $a^2$-π*$b^2$. (Figure S7) The current collector cannot be the whole bottom area anymore. The center is replaced by a circle electrolyte with radius *b*. (The grey part in the Figure S7) The current collector is smaller than an intact one on a homogenous electrode. Thus, for cell whose vascular channels are defined on the LCO electrode, the 1C current density would be:

$$i_{1C} = \frac{a^2}{a^2 - \pi b^2} \cdot \frac{Q}{3600\,[s]}$$

After calculation, the real areal capacity versus voltage would be calculated for charging time $t$:

$$Q_{real} = \frac{t\,[s]}{3600\,[s]} \cdot i_{1C} \cdot \frac{a^2 - \pi b^2}{a^2} \cdot C$$

Where C is the charging rate.

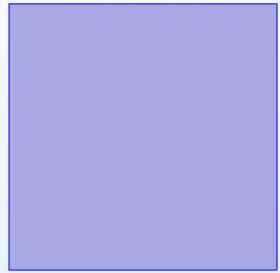

Figure S6 Current collector on homogeneous LCO electrode. The current can be exerted on the whole interface (blue shadow).

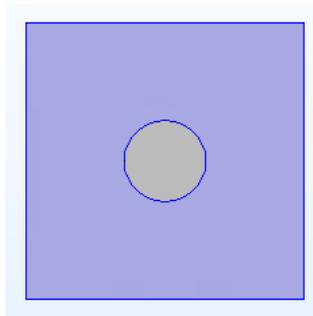

Figure S7. Current collector on LCO electrode with channels. The blue shadow is a solid phase, which can be exerted with the current, while the grey shadow (electrolyte) can not.

**Note 4. Tortuosity calculation method**

According to the previous literature, the tortuosity of porous electrodes is defined by the

following formula:

$$\tau = \varepsilon \frac{D_0}{D_{eff}}$$

$D_0$ is the intrinsic diffusivity of $Li^+$ in the electrolyte and $D_{eff}$ is the effective diffusivity of $Li^+$ in the porous electrode-electrolyte phase. Based on the Bruggeman exponent, the effective diffusivity is calculated by the local average porosity of porous electrode: [5]

$$D_{eff} = \varepsilon^{\frac{3}{2}} D_0$$

It shows that the tortuosity is dependent on the local average porosity at a certain xy-plane $\varepsilon$ of the electrode. For the electrode with vascular structures, the porosity in one electrode are different for the layers with a different number of branches. Therefore, the tortuosity of such kind of electrode would be a function of position in the electrode. To obtain the overall effective tortuosity of the vascularized porous electrode, we first need to calculate the effective diffusivity of each layer with the different number of branches based on the local average porosity (or density). The local effective diffusivity calculation of different layer can be calculated by the below equations:

$$\rho_{solid,n} = \frac{[a^2 - num_n \cdot \pi \cdot (\alpha_n \cdot b)^2 - \pi \cdot b^2] \cdot \rho_{vascular\text{-}electrode}}{a^2}$$

$$\rho_{electrode,n} = \frac{\rho_{solid,n}}{\rho_{vascular\text{-}electrode}}$$

$$D_{eff,vascular} = (1 - \rho_{vascular\text{-}electrode})^{\frac{3}{2}} \cdot D_0$$

$$D_{eff,n} = \rho_{electrode,n} \cdot D_{eff,vascular} + (1 - \rho_{electrode,n}) \cdot D_0$$

The subscript n refers to the label of each layer (1$^{st}$ layer, 2$^{nd}$ layer, etc.). $\rho_{vascular\text{-}electrode}$ is the density of the porous electrode region when having vascularized channels, which is equal to 0.7. $\rho_{solid}$ means the solid phase volume fraction which is calculated by the unit cell size *a*, central channel radius *b*, ratio of branch's radius to central channel's radius *alpha*, and the number of branches in the layer *num*. $\rho_{electrode}$ is the porous electrode region's volume fraction. The effective diffusivity in the porous electrode region is equal to $D_{eff,\,vascular}$ (calculated by the density of porous electrode phase $\rho_{vascular\text{-}electrode}$), and the rest region is the channels, whose diffusivity is equal to $D_0$. The overall effective diffusivity in each layer is calculated by weighted-average summing up the effective diffusivity in two regions (porous electrode and channels). Given the effective diffusivity of each layer, the tortuosity of each layer could also be written as below:

$$\tau_n = (1 - \rho_{solid,n}) \cdot \frac{D_0}{D_{eff,n}}$$

We could also express the volume fraction $\rho_{layer}$ of each layer. Since the whole unit cell has the same cross-sectional area, the volume fraction would degrade to the thickness fraction of each layer:

$$\rho_{layer,n} = \frac{h_n}{h}$$

Thus, the overall tortuosity of the whole electrode could be calculated by weighted-average summing up the tortuosity of each layer n:

$$\tau = \sum_n \tau_n \cdot \rho_{layer,n}$$

Note that the tortuosity of the homogenous electrode could be calculated by its overall porosity ($\varepsilon_{homo} = 0.4$):

$$D_{eff,homo} = \varepsilon_{homo}^{\frac{3}{2}} \cdot D_0$$

$$\tau_{homo} = \varepsilon_{homo} \cdot \frac{D_0}{D_{eff,homo}}$$

## Note 5. Derivation of penetration depth as a function of porosity

So far, we have observed that graded porosity (or gradient-distributed branches in our vascular simulation) can improve battery performance. However, few research focused on the basic physical principles causing this phenomenon. To study the relationship between porosity and Li-ion battery charging capacity, we need to go into the basic transportation theory in the porous electrode. Previously, John Newman has solved the partial differential equation of particle transportation in the Li-ion battery porous electrode. [1] However, he only solved this for the discharging process. Here, to study the charging process, we change the boundary condition and apply the linear Taylor approximation to the PDE group developed by John Newman. We modify the penetration depth given by John Newman to get the relation between penetration depth and porosity in the charging process.

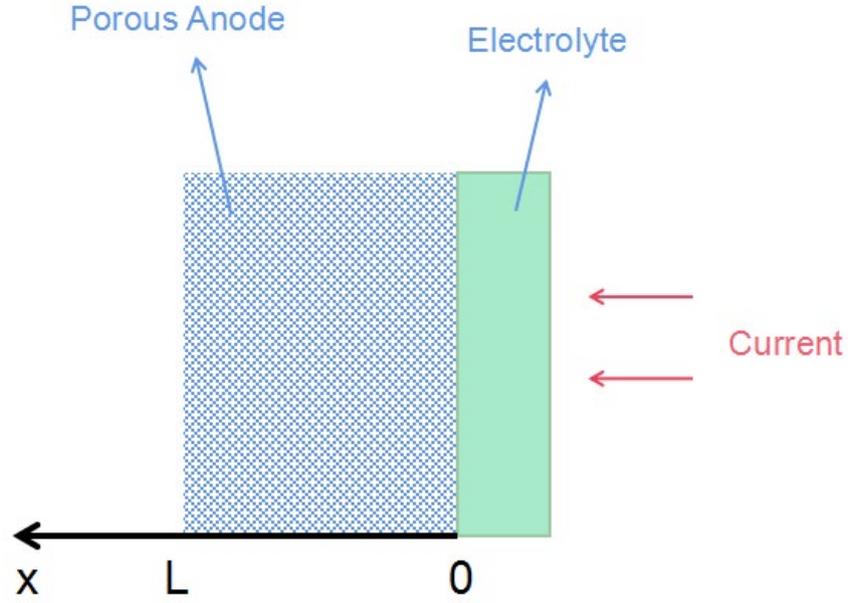

Figure S8. Illustration of porous electrode charging model and its coordinate.

According to John Newman's theory for 1D porous electrode, the 3D partial differential equation groups could be simplified to 1D form:

$$i_2 = -\kappa \frac{d\Phi_2}{dx} \qquad (1)$$

$$i_1 = -\sigma \frac{d\Phi_1}{dx} \qquad (2)$$

$$\frac{di_1}{dx} + \frac{di_2}{dx} = 0 \qquad (3)$$

$$\frac{di_2}{dx} = a \cdot f(\Phi_1 - \Phi_2) \qquad (4)$$

We define the Cartesian coordinate along the anode, starting from the separator as in Figure S9. Then we have the boundary condition as below:

$$i_2 = I, \ i_1 = 0, \ \text{at } x = 0$$

$$i_2 = 0, \text{ at } x = L$$

In the equations above, σ is the electric conductivity of the matrix materials; $i_1$ is the superficial current density in the matrix phase. $i_2$ is the current in the pores phase. κ is the effective conductivity of the pore solution which could be expressed as:

$$\kappa = \varepsilon F^2 \sum_i z_i^2 u_i c_i$$

Where F is the Faraday constant, $z_i$ is the charge number of species i, $u_i$ is the ionic mobility and $c_i$ is the superficial average concentration in species i.

Since the polarization equation in 3D model which describe the charge transfer from solid to pores phase could be written as Butler-Volmer equation as below:

$$\nabla \cdot i_2 = a i_0 \left[ \exp\left(\frac{\alpha F \eta_s}{RT}\right) - \exp\left(\frac{(1-\alpha)F\eta_s}{RT}\right) \right] \quad (5)$$

where a is the specific area which is in the unit of (1/m), $\eta_s$ is the overpotential and could be expressed as $\eta_s = \Phi_1 - \Phi_2$ when $\Phi_2$ is measured by the reference electrode with the same materials as working electrode, α is the transfer coefficient.

We can further simplify the polarization equation eq.(6) by using Talyor linear approximation. In addition, the process we consider is charging. Hence the eq.(4) would become:

$$\frac{di_2}{dx} = c(\Phi_1 - \Phi_2)$$

where

$$c = \frac{a i_0 F}{RT}$$

Such equation could be transferred to the form below, by applying boundary conditions

$$\frac{d^2 i_2}{dx^2} = c\left(\frac{d\Phi_1}{dx} - \frac{d\Phi_2}{dx}\right)$$
$$= c\left[-\frac{I}{\sigma} + i_2\left(\frac{1}{\kappa} + \frac{1}{\sigma}\right)\right] \quad (6)$$

To solve this, we use variable substitution by defining substitutive variables:

$$j = \frac{i_2}{I},\ y = \frac{x}{L},\ v = L\sqrt{c\frac{\kappa+\sigma}{\kappa\sigma}},\ \delta = \frac{cL^2}{\sigma}$$

For the left-hand side of eq. (6)

$$\frac{d^2 i_2}{dx^2} = \frac{d}{dx}\left(\frac{di_2}{dx}\right)$$

$$= \frac{d}{dx}\left(\frac{d(jI)}{dy}\cdot\frac{dy}{dx}\right)$$

$$= \frac{d}{dx}\left(\frac{I}{L}\cdot\frac{dj}{dy}\right)$$

$$= \frac{I}{L^2}\frac{d^2 j}{dy^2}$$

Therefore, the equation could be transferred to the below form

$$\frac{d^2 j}{dx^2} = -\delta + v^2 j$$

However, the expressions above didn't show the relationship with porosity explicitly. In order to demonstrate the porosity influence, we need to substitute the specific area $a$ into porosity $\varepsilon$. For sphere electrode particle, the specific area could be written in terms of mean particle radius r

$$a = \frac{3(1-\varepsilon)}{r}$$

Thus, the constant $v$ could be expressed as

$$v = L\sqrt{\frac{3(1-\varepsilon)i_0 F}{rRT}\frac{\kappa+\sigma}{\kappa\sigma}}$$

Now, we can introduce the reaction penetration depth to describe how thick the electrode could be utilized effectively. With all the solution and equation above, we can modify the penetration depth proposed by John Newman [1]

$$\frac{L}{v} = \sqrt{\frac{rRT}{3(1-\varepsilon)i_0 F}\frac{\kappa\sigma}{\kappa+\sigma}} = a_1\sqrt{\frac{1}{1-\varepsilon}}$$

where $a_1$ is a constant.

From the equation, we can easily see that the reaction could penetrate deeper by increasing the electrode porosity, which will improve the specific capacity during charging with a high C-rate. The improvement of penetration depth by increasing average porosity is plotted in Figure S9.

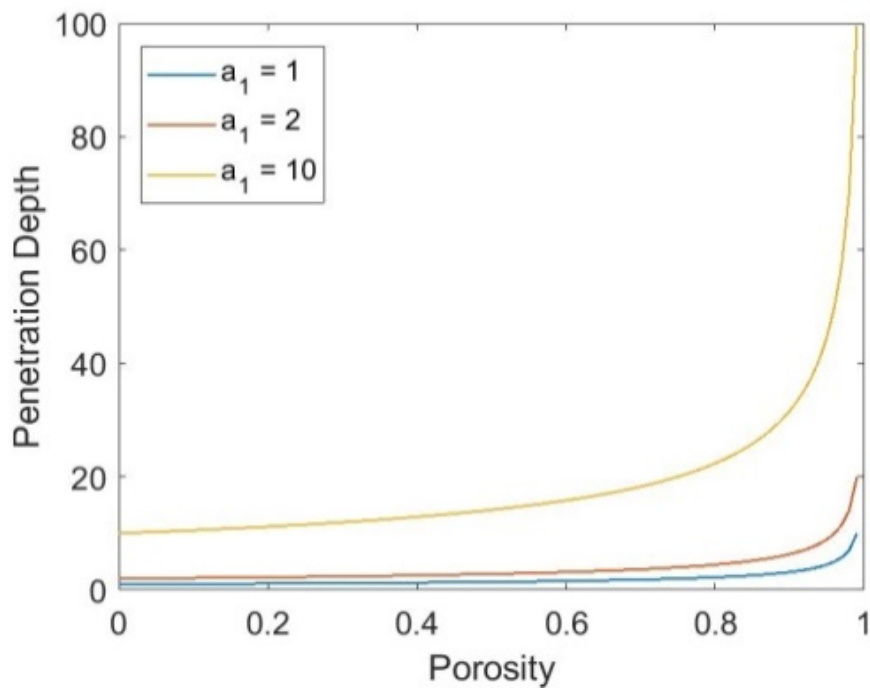

Figure S9. Penetration depth as the function of porosity.

**Note 6. Deep learning and inverse design method**

To predict the capacity of different geometry, we used a fully connected layer artificial neural network (ANN) as a prediction model. The input of the neural network includes the geometric parameters of vascular electrode structure with the corresponding local average porosity profile of each vascular section and the C-rate. The output of the neural network is the charging curve composed of 20 points, which are the corresponding capacities evenly spaced versus the voltage from 2.5 V (initial potential) and 4.3 V (cut-off potential), shown in Figure S10.

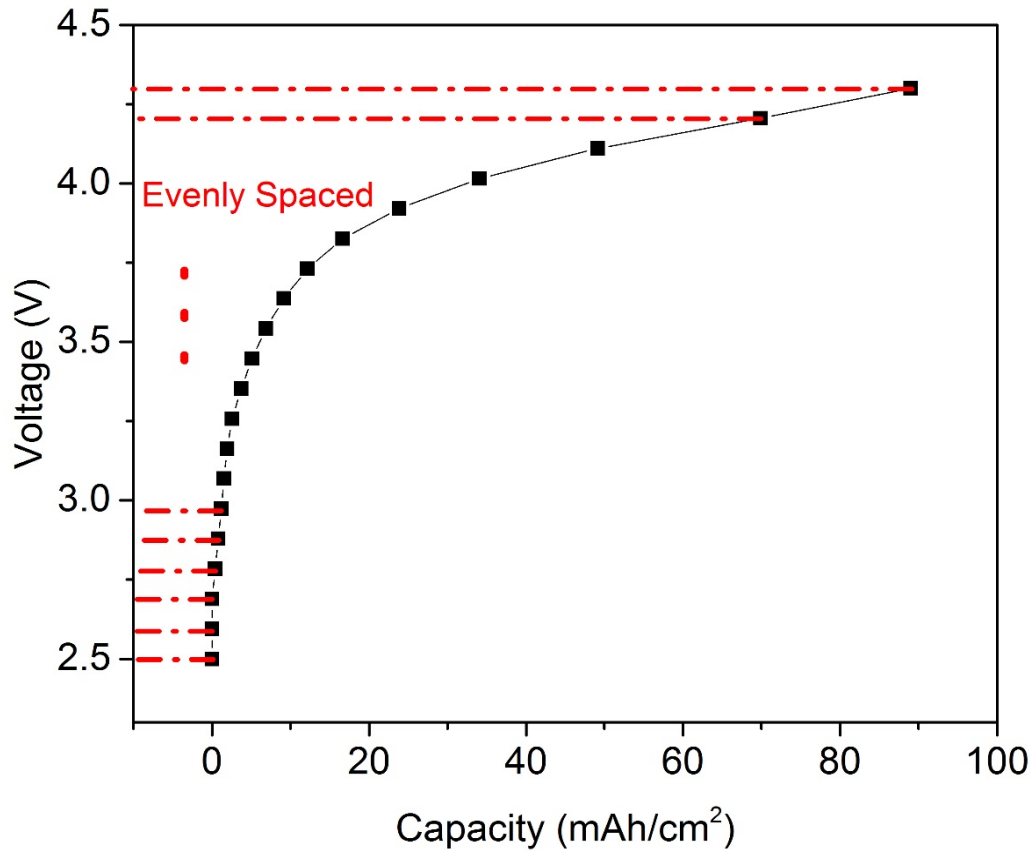

Figure S10. Demonstration how 20 evenly spaced data points are picked from COMSOL charging curves.

The neural network computation graph would be constructed as 5 hidden layers with 150, 300, 300, 150, 50 neurons, respectively. The input is a 14*1 vector containing geometric parameters of each vascular electrode structure, average porosity in each section of vascularized electrode ($\varepsilon_A$, $\varepsilon_B$ and $\varepsilon_C$) shown in Figure S12, and the corresponding C-rate. The output is a 20*1 vector corresponding to the charging capacities with respect to the voltage of each vascular structure. After prediction, 20 capacity points are re-arranged to a 20*2 matrix where each row contains the capacity and its corresponding voltage. The matrix is utilized to plot the charge curve. The conceptual diagram of neural network's structure is shown in the Figure S11. The activation function of each layer is the ReLU function. For the process of updating weights, we used the ADAM optimizer. The PyTorch built-in loss function is used to determine the Mean Square Error (MSE) loss. To train the fully connected layer artificial neural network, 4611 charging curves (92220 capacity data points) are generated from finite element simulation in COMSOL. The vascular structure geometries are randomly generated from MATLAB.

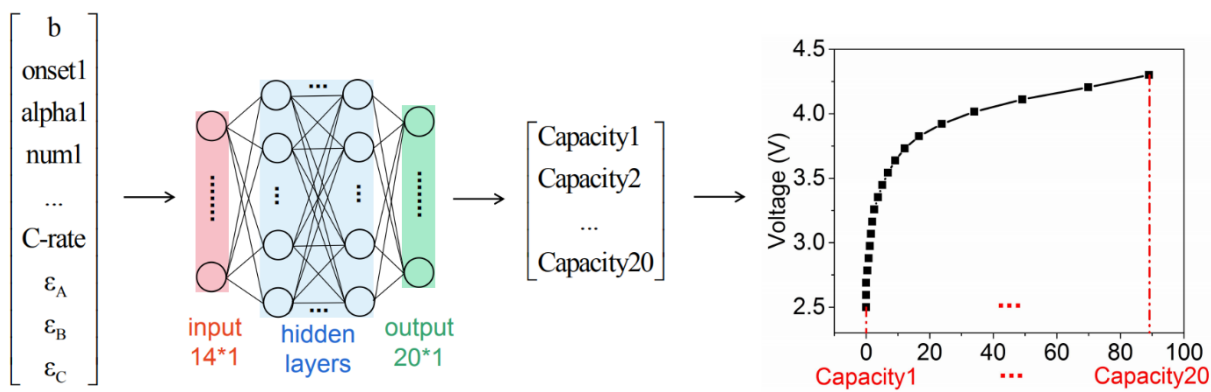

Figure S11. Schematic diagram of neural network and the data structures for its input and output.

We conduct the following calculation to compare the time consumption for the traditional finite element method versus our statistical deep learning method. The training data set we choose contained 4611 geometries and last 20 days to finish their corresponding finite element computation. Afterwards, 100 different neural networks were trained to conduct the bagging ensemble. This process takes about 1 hour. Thus, the time we built the total charging curves library is about 20*24=480 hours. If we used the finite element method to generate the charging curves for all 389514 geometries, it would take 40548 hours (about 4.62 years) to finish all computations by assuming each computation costs the same time as generating the training dataset.

Since the traditional neural network deep learning approach can only generate a point estimation of the prediction, one cannot get the uncertainty of the results. We choose the bootstrap aggregating (bagging) ensemble algorithm to make our prediction results more stable and accurate. 100 separated neural networks were trained simultaneously with different subsets randomly chosen from the overall training data. The final prediction is the mean of those predictions from 100 trained neural networks. With the bagging method, we can not only improve the predicting accuracy but also provide the corresponding variance of capacity in our searching library. Likewise, other useful electrochemical properties such as energy density, power density, and average voltage were also computed with the bagging method.

In the inverse searching process, we first searched and sorted concerning the vascular branch number configuration and found the top 3 cases. In each optimal vascular branch number configuration, we did the conditional search again and provided the top 2 geometry parameters that match the searching condition the best. These 6 geometries are the exact full geometry parameters, and their packages guiding customers on how to fabricate the vascular structures (how large the central channel and branch channels are, where the onset nodes should be).

As for the definition of objective electrochemical criteria, the charging capacity is defined as the capacity when the charging curve hits the cut-off voltage. Energy density is defined as the integral of charging curves from initial time to the end of charging time. The average voltage is the energy density divided by the charging capacity. Power density is defined as the production of average voltage and current density under each charging condition.

| A | num1 | num2 | num3 | b | onset1 | alpha1 | onset2 | alpha2 | onset3 | alpha3 | Capacity | Power Density |
|---|---|---|---|---|---|---|---|---|---|---|---|---|
| #1 | 0 | 0 | 4 | 0.01 | – | – | – | – | 0.14 | 1 | 122.6±0.007 | 1962.97±0.078 |
| #2 | 0 | 0 | 4 | 0.01 | – | – | – | – | 0.12 | 1 | 121.9±0.005 | 1962.12±0.091 |
| #1 | 0 | 2 | 4 | 0.01 | – | – | 0.12 | 1 | 0.16 | 1 | 120.1±0.012 | 1961.20±0.092 |
| #2 | 0 | 2 | 4 | 0.01 | – | – | 0.1 | 1 | 0.14 | 1 | 119.9±0.007 | 1960.77±0.083 |
| #1 | 2 | 0 | 4 | 0.01 | 0.06 | 1 | 0.1 | – | 0.14 | 1 | 119.9±0.017 | 1961.52±0.114 |
| #2 | 2 | 0 | 4 | 0.01 | 0.04 | 1 | 0.08 | – | 0.14 | 1 | 119.9±0.015 | 1961.40±0.114 |
| B | num1 | num2 | num3 | b | onset1 | alpha1 | onset2 | alpha2 | onset3 | alpha3 | Energy Density | Power Density |
| #1 | 0 | 2 | 4 | 0.005 | – | – | 0.12 | 1 | 0.14 | 1 | 28.56±0.014 | 3984.46±0.39 |
| #2 | 0 | 2 | 4 | 0.005 | – | – | 0.12 | 1 | 0.16 | 1 | 28.55±0.010 | 3984.52±0.43 |
| #1 | 0 | 0 | 4 | 0.006 | – | – | – | – | 0.12 | 1 | 28.58±0.008 | 3984.94±0.35 |
| #2 | 0 | 0 | 4 | 0.005 | – | – | – | – | 0.12 | 1 | 28.74±0.009 | 3984.97±0.38 |
| #1 | 0 | 4 | 4 | 0.005 | – | – | 0.12 | 1 | 0.14 | 1 | 28.32±0.012 | 3985.14±0.55 |
| #2 | 0 | 4 | 4 | 0.005 | – | – | 0.14 | 1 | 0.16 | 1 | 28.41±0.013 | 3985.23±0.54 |

Table S4. Optimized vascular structures' demonstration under different criteria. The units in this sheet: b, onset: mm, Capacity: mAh/cm$^2$, Energy Density: Wh/m$^2$, Power Density: W/m$^2$.

**Note 7. Definition of average porosity distribution's classification**

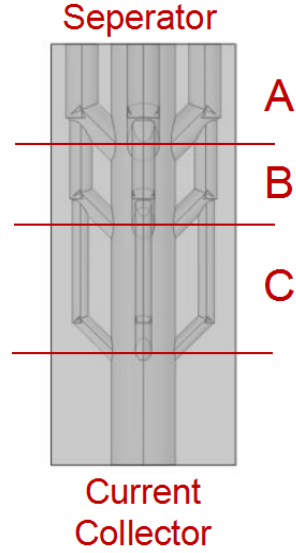

Figure S12. Average porosity distribution analysis schematics. The vascular channel is separated into 3 regions A, B, and C.

As mentioned in the manuscript, vascular structures introduce the average porosity at a certain xy-plane distribution along the electrode (z-axis direction), which mainly contributes to the difference of performance of each vascular structure. To analyze how the vascular structure changes the average porosity distribution and how it impacts the charging capacity, we define 13 kinds of porosity distribution profiles. As shown in Figure S12, we divide the vascular channel into 3 sections since there are at most 3 onset nodes. In each section, the branch number remains the same since this section is only contributed by its corresponding onset node. For instance, the branch number in section A is only determined by the *num1*. Section B and C are also similar. Therefore, in each section, the average porosity at a certain xy-plane could be calculated by the following equation:

$$\varepsilon_n = 1 - \frac{\left[a^2 - num_n \cdot \pi \cdot (\alpha_n \cdot b)^2 - \pi \cdot b^2\right] \cdot \rho_{\text{porous-electrode}}}{a^2}$$

Where *a* is unit cell size, $num_n$ is the number of branches grown on the onset node, $\alpha_n$ is the ratio between central channel radius b and the branch radius. $\rho_{\text{porous-electrode}}$ is the solid phase density of porous electrode (equal to 0.7 when having vascular structure). Thus, we define 13 kinds of porosity distribution profiles according to the porosity difference between adjacent sections. Among the profiles, 'h', 'e', 'l', represents 'higher', 'equivalent' and 'lower', respectively. If $\varepsilon_A > \varepsilon_B > \varepsilon_C$, then we name this kind of profile as 'hh', meaning 'higher and higher'. Likewise, all porosity distribution profiles are listed in table S5 below.

| Porosity distribution profile | Nomenclature |
|---|---|
| $\varepsilon_A > \varepsilon_B > \varepsilon_C$ | hh |
| $\varepsilon_A > \varepsilon_B = \varepsilon_C$ | he |
| $\varepsilon_A > \varepsilon_B < \varepsilon_C$ and $\varepsilon_A > \varepsilon_C$ | hl-good |
| $\varepsilon_A = \varepsilon_B > \varepsilon_C$ | eh |
| $\varepsilon_A < \varepsilon_B > \varepsilon_C$ and $\varepsilon_A > \varepsilon_C$ | lh-good |
| $\varepsilon_A > \varepsilon_B < \varepsilon_C$ and $\varepsilon_A = \varepsilon_C$ | hl-e |
| $\varepsilon_A = \varepsilon_B = \varepsilon_C$ | ee |
| $\varepsilon_A = \varepsilon_B < \varepsilon_C$ | el |
| $\varepsilon_A > \varepsilon_B < \varepsilon_C$ and $\varepsilon_A < \varepsilon_C$ | hl-bad |
| $\varepsilon_A < \varepsilon_B > \varepsilon_C$ and $\varepsilon_A = \varepsilon_C$ | lh-e |
| $\varepsilon_A < \varepsilon_B > \varepsilon_C$ and $\varepsilon_A < \varepsilon_C$ | lh-bad |
| $\varepsilon_A < \varepsilon_B = \varepsilon_C$ | le |
| $\varepsilon_A < \varepsilon_B < \varepsilon_C$ | ll |

Table S5. Porosity distribution profiles and their nomenclatures.

As mentioned earlier, porosity distribution impacts the performance of the battery significantly. For vasculatures, the porosity distribution is determined by the number, position, and radius of branches. Although the improvement by vasculature is obvious, the optimal porosity distribution profile is hard to determine based on the limited physical intuition, which may also change dramatically with different kinds of battery. To analyze the influence of porosity distribution profiles on the charging performance, we define 13 porosity distribution profiles (table S5) and count the proportions of various porosity distribution profiles in different vascular structures. All 389514 geometries are sorted into 24 different types based on the branch number configurations (the number of branches on each onset point). The relative global performance of different vascular branch number configurations is compared by the counting score's method (Supporting Information Note 8). In Figure S13, the x-axis represents different vascular branch number configurations, and the order is ranked from the best to the worst charging capacity under 5C (i.e., from 004 to 440). The y-axis represents different types of porosity distribution profiles. The order of porosity profiles is ranked from the best to the worst based on the "high porosity near separator-electrode region" rule (from "hh" to "ll"). The vascular branch number configurations with better performance tend to have a decreasing porosity distribution from the separator-electrode interface to the current collector ('hh', 'he', and 'hl-good'). Configurations like '440' and '420' tend to have worse performance because their porosity is completely contrary to the decreasing porosity distribution. With this porosity distribution analytical map, locating the optimized vascular branch number configurations from our model is more crucial for understanding the physical principles and guiding the fabrication. The optimized configurations are labeled in Figure S13 as red squares. It is obvious that all optimized vascular geometries are located on the upper left corner of the heat map and obey the "high porosity near separator-electrode region" rule (check the porosity profile definition in table S5). This porosity profile analysis gives readers the physical intuition about how vascular structures enhance the battery's high-rate performance.

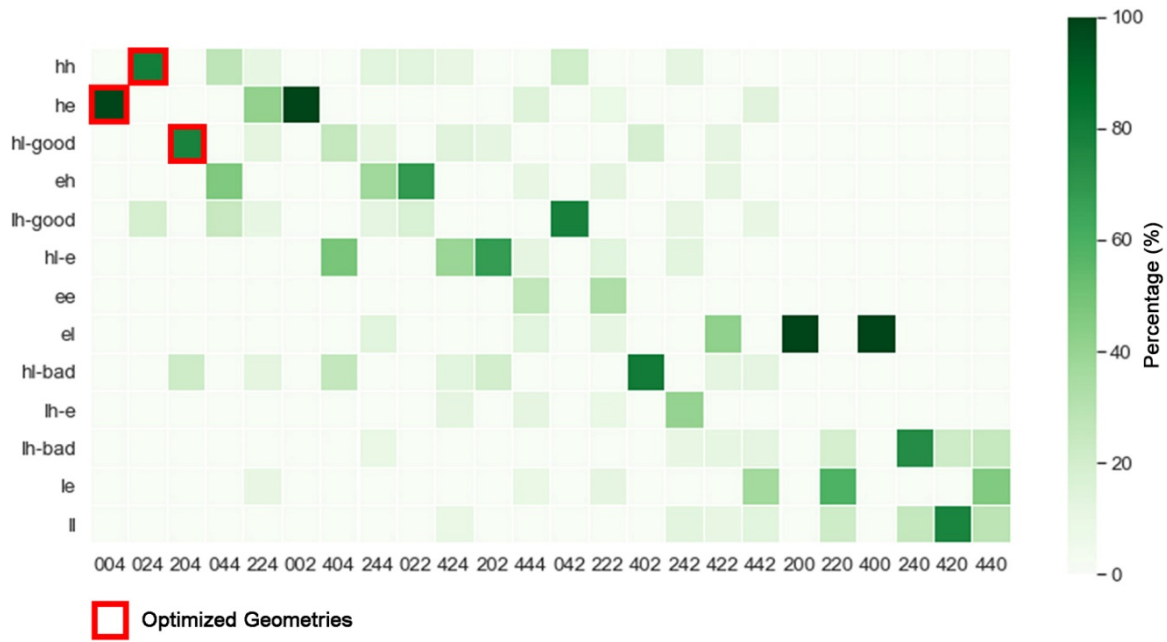

Figure S13. The porosity profiles analysis for all vascular structures and the location of optimized geometries in table S4 are labeled as a red square in the heat map. The color bar

demonstrates the percentage of each porosity profile in different vascular branch number configurations. Here, "h", "e", and "l" are the abbreviations for "higher", "equal", and "lower", respectively. They represent the relationship between each section of the electrode from the electrode-separator interface to the current collector. For example, "hh" means the porosity of the first section (close to the separator) is higher than the one of the second section and the porosity of the second section is higher than the one of the third section (close to the current collector). The rest is similar. For some profiles like "hl", a more detailed classification is introduced based on the global porosity trend. If the porosity follows the "high porosity near separator-electrode region" rule, we name them as "good", otherwise "bad". If the porosity near the separator is the same as the one near the current collector, then the profile would be named as "e". (Table S5)

**Note 8. Counting scores' method**

We also classify all 389514 geometries into 24 different categories based on the number of branches on each onset node. The nomenclature format is: (num1, num2, num3). To demonstrate the global performance of each vascular branch's number configuration, we design a simple scoring system. Firstly, we rank the parameters based on the chosen optimization criteria (from large to small capacities, from high to low energy densities). Then, we label the score on each parameter: the $n^{th}$ place parameter would win (389514 - n) point for its corresponding branches number configuration. Lastly, the scores are divided by the numbers of the parameters to get the "average score" of each branch's number configuration. With the scoring system, we can determine which branch number configuration has better performance.

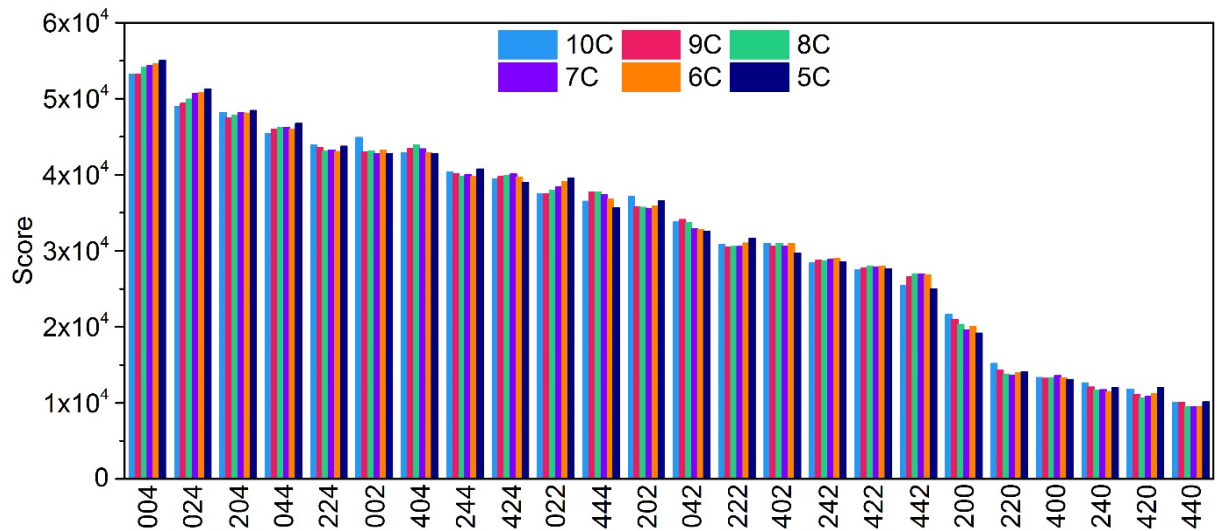

Figure S14. Charging capacity scores comparation for all vascular branches number configurations

**Note 9. Global sensitivity analysis**

Sobol indices analysis is used to analyze the contribution of each geometric factor to the charging capacity. The larger the index, the more significant its corresponding geometric factor. [6, 7] We use a well-known sensitivity analysis package SALib to calculate Sobol indices of 10 geometry parameters under different charging rates.

It is important to note that the summation of all the Sobol indices of each charging rate are all slightly larger than 1 (from 1.03 to 1.06), this could be explained by the strong correlation between each parameter. The strong correlation between input variables would cause the slight inaccuracy of global sensitivity analysis results. [8] The correlation relationship of 10 geometric parameters is shown in Figure S13. The correlation between all onset nodes' positions is the most noticeable. Other parameters also correlate with each other. Hence, it is not surprising to obtain the Sobol indices results whose summations of different C-rate are slightly larger than 1.

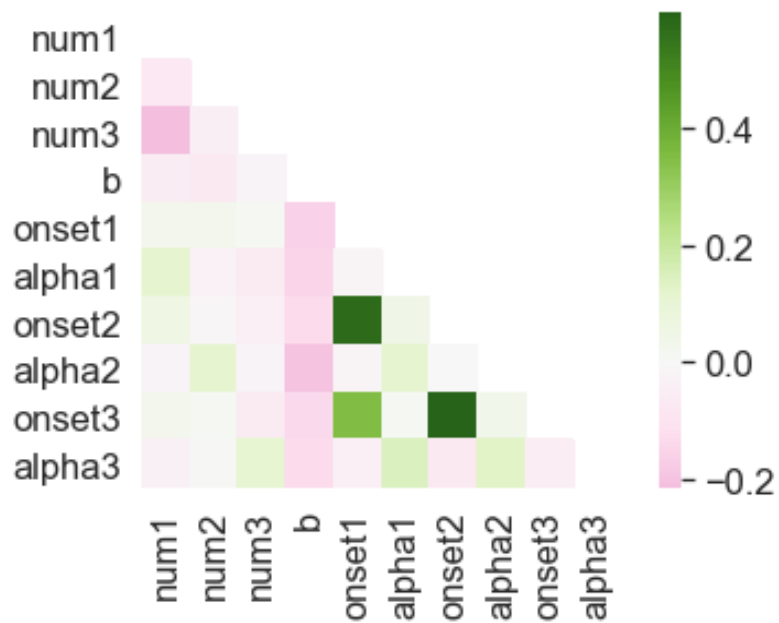

Figure S15. Correlation matrix of geometric parameters.

**Note 10. Supplementary data**

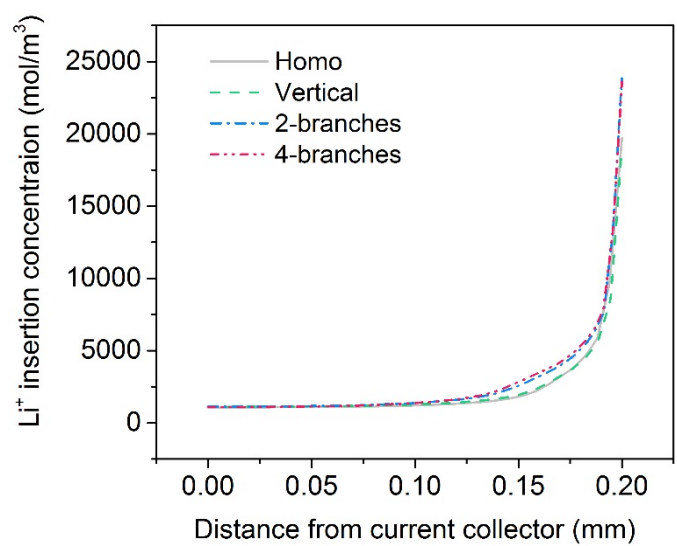

Figure S16. Li-ion insertion depletion profile.

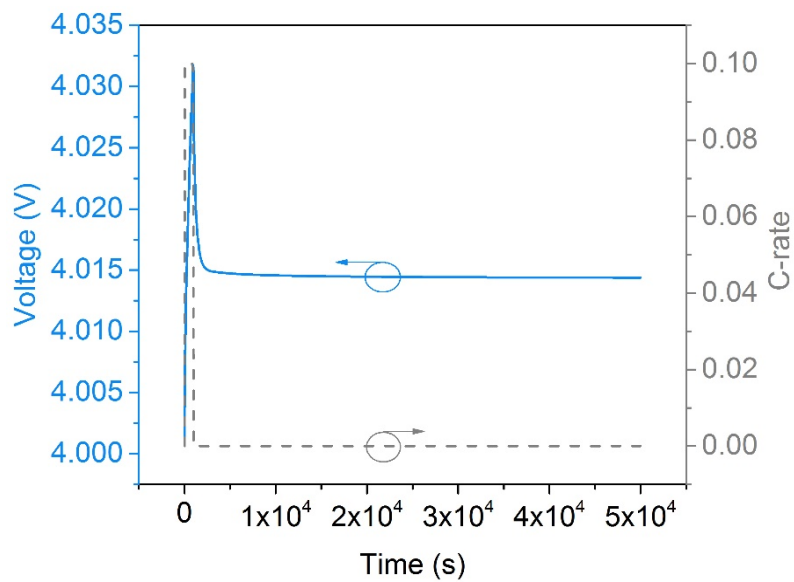

Figure S17. Voltage and current profiles of the polarization-depolarization process.

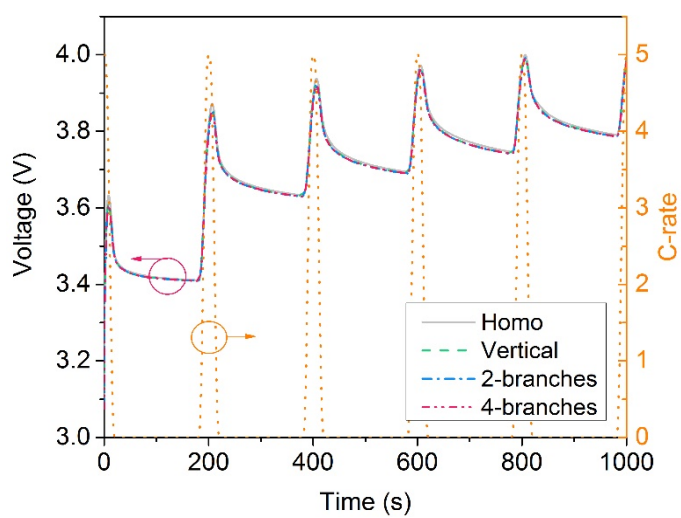

Figure S18. High C-rate pulse charge voltage and current profiles.

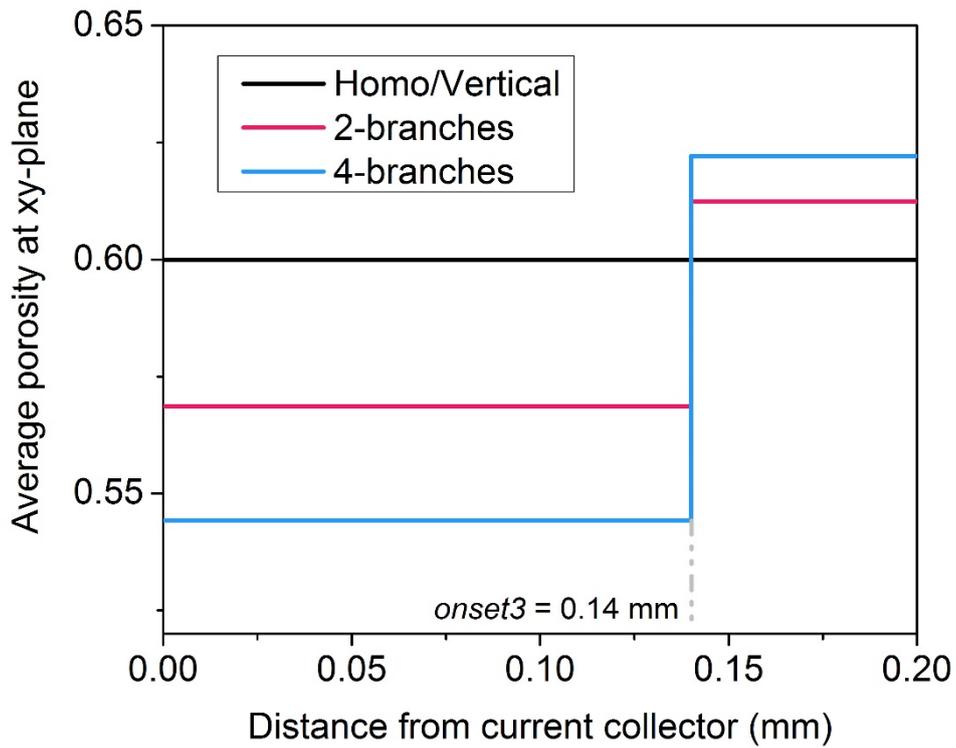

Figure S19. Average porosity distribution along the electrode *z*-direction calculated by the method defined in Supporting Information Note 7.

| Parameter | 5C | 6C | 7C | 8C | 9C | 10C |
|---|---|---|---|---|---|---|
| num1 | 0 | 0 | 4 | 2 | 4 | 2 |
| num2 | 0 | 4 | 4 | 4 | 2 | 2 |
| num3 | 4 | 2 | 0 | 4 | 0 | 0 |
| b | 0.005 | 0.009 | 0.016 | 0.007 | 0.005 | 0.005 |
| onset1 | - | - | 0.02 | 0.04 | 0.02 | 0.02 |
| alpha1 | - | - | 0.5 | 0.5 | 0.5 | 0.5 |
| onset2 | - | 0.14 | 0.06 | 0.06 | 0.06 | 0.04 |
| alpha2 | - | 0.5 | 0.5 | 1 | 0.5 | 1 |
| onset3 | 0.12 | 0.18 | 0.14 | 0.14 | 0.08 | 0.12 |
| alpha3 | 0.5 | 0.5 | - | 0.5 | - | - |

Table S6. Geometric parameters for validation test examples.

**Reference:**

[1] Newman, J., and Thomas-Alyea, K. E. (2012) Electrochemical systems. John Wiley & Sons.

[2] Kandler, S. and Wang, C.Y. (2006) Power and thermal characterization of a lithium-ion battery pack for hybrid-electric vehicles. Journal of power sources 160, 662-673.

[3] Rui, Z., Liu, L. and Ma, F. (2020) Cathode Chemistries and Electrode Parameters Affecting the Fast Charging Performance of Li-Ion Batteries. Journal of Electrochemical Energy Conversion and Storage 17.

[4] Bae, C.J., Erdonmez, C.K., Halloran, J.W., and Chiang, Y.M. (2013). Design of battery electrodes with dual-scale porosity to minimize tortuosity and maximize performance. Adv Mater 25, 1254-1258.

[5] Cooper, S.J., Eastwood, D.S., Gelb, J., Damblanc, G., Brett, D.J.L., Bradley, R.S., Withers, P.J., Lee, P.D., Marquis, A.J., Brandon, N.P. and Shearing, P.R. (2014) Image based modelling of microstructural heterogeneity in LiFePO4 electrodes for Li-ion batteries. Journal of Power Sources 247, 1033-1039.

[6] Sobol, I.M. (2001). Global sensitivity indices for nonlinear mathematical models and their Monte Carlo estimates. Mathematics and Computers in Simulation 55, 271–280.

[7] Herman, J., and Usher, W. (2017). SALib: An open-source Python library for Sensitivity Analysis. The Journal of Open Source Software 2.

[8] Do, N.C., and Razavi, S. (2020). Correlation Effects? A Major but Often Neglected Component in Sensitivity and Uncertainty Analysis. Water Resources Research 56.